\documentclass[traditabstract]{aa}
\usepackage{txfonts}

\usepackage{graphicx}
\usepackage{natbib}
\usepackage{color} 
\newcommand{\nc}{\newcommand}
\newcommand{\cmsq}{\mbox{cm$^{-2}$}}
\newcommand{\kms}{\mbox{km~s$^{-1}$}}

\newcommand{\cplus}{\mbox{C$^+$}}
\newcommand{\HII}{H {\sc ii}}
\nc{\Msun}{\ensuremath{\mathrm{M}_\odot}}
\nc{\lsun}{\ensuremath{\mathrm{L}_\odot}}
\nc{\cmcub}{\mbox{cm$^{-3}$}}
\nc{\Kkms}{\mbox{K~km/s}}
\nc{\CeiO}{C$^{18}$O}
\nc{\thCO}{$^{13}$CO}
\nc{\vlsr}{\mbox{$\upsilon_{\rm LSR}$}}
\nc\micron{\mbox{$\mu$m}}
\newcommand{\thCII}{[$^{13}$C {\sc ii}]}
\newcommand{\CII}{[C {\sc ii}]}
\newcommand{\OI}{[O {\sc i}]}
\newcommand\arcdeg{\mbox{$^\circ$}}%
\newcommand\phn{\phantom{0}}%
\newcommand\tkin{$T_{\rm kin}$}%

\def\ptsec{$''\mskip-7.6mu.\,$}
\def\psec{$^s\mskip-7.6mu.\,$}

\newcommand{\formaldehyde}{H$_2$CO}
\newcommand{\formaldehydethtwo}{H$_2$CO(3$_{03}$--2$_{02}$)}

\usepackage{txfonts}
\begin{document}

   \title{Opening the Treasure Chest in Carina}

   \subtitle{}

   \author{B. Mookerjea \inst{1} 
   \and G. Sandell \inst{2} 
   \and R. G\"usten \inst{3} 
   \and D. Riquelme \inst{3}
   \and H. Wiesemeyer \inst{3}
   \and E. Chambers \inst{4}
   }

   \institute{Tata Institute of Fundamental Research, Homi Bhabha Road,
Mumbai 400005, India
              \email{bhaswati@tifr.res.in}
         \and
         Institute for Astronomy, University of Hawaii,  640 N. Aohoku Place, Hilo, HI 96720, USA
         \and
Max Planck Institut f\"ur Radioastronomie, Auf dem H\"ugel 69, 53121 Bonn, Germany     
\and 
USRA/SOFIA, NASA Ames Research Center, Mail Stop 232-12, Building N232, P.O. Box 1, Moffett Field, CA 94035-0001, USA
             }


 
\abstract{
We have mapped the G287.84-0.82 cometary globule (with the Treasure
Chest cluster embedded in it) in the South Pillars region of Carina (i)
in \CII, 63\,\micron\ \OI\ , and CO(11--10) using the heterodyne
receiver array upGREAT on SOFIA and (ii) in $J$=2--1 transitions of CO,
\thCO, \CeiO\ and $J$=3--2 transitions of H$_2$CO using the APEX
telescope in Chile. We use these data to probe the morphology,
kinematics, and physical conditions of the molecular gas and the photon
dominated regions (PDRs) in G287.84-0.82. The velocity-resolved
observations of \CII\ and \OI\ suggest that the overall structure of
the pillar (with red-shifted photo evaporating tails) is consistent
with the effect of FUV radiation and winds from $\eta$\,Car and O stars
in Trumpler 16. The gas in the head of the pillar is strongly
influenced by the embedded cluster, whose brightest member is an  O9.5
V star, CPD -59\arcdeg 2661. The emission of the \CII\ and \OI\ lines
peak at a position close to the embedded star, while all the other
tracers peak at another position lying to the north-east consistent
with gas being compressed by the expanding PDR created by the embedded
cluster. The molecular gas inside the globule is probed with the
$J$=2--1 transitions of CO and isotopologues as well as H$_2$CO,  and
analyzed using a non-LTE model (escape-probability approach), while we
use PDR models to derive the physical conditions of the PDR. We
identify at least two PDR gas components; the diffuse part
($\sim$10$^4$\,\cmcub) is traced by  \CII, while the dense
($n\sim$2--8\,10$^5$\,\cmcub) part is traced by \CII, \OI, CO(11--10).
Using the the $F$=2--1 transition of \thCII\ detected at 50 positions
in the region, we derive optical depths (0.9--5), excitation
temperatures of  \CII\ (80--255\,K), and $N$(\cplus) of 0.3--1$\times
10^{19}$\,\cmsq. The total mass of the globule is $\sim$ 1,000 \Msun,
about half of which is traced by \CII. The dense PDR gas has a thermal
pressure of $10^7$--$10^8$\,K\,\cmcub, which is similar to the values
observed in other regions.  }


   \maketitle
%

\section{Introduction}

Some of the most spectacular structures in the molecular interstellar
medium (ISM) are observed in the vicinity of hot O/B stars or
associations in the forms of cometary globules and pillars. The cometary
globules arise when the expanding \HII\ region due to the O/B star
overruns and compresses an isolated nearby cloud causing it to collapse
gravitationally. As the cloud contracts the internal thermal pressure
increases causing the cloud to re-expand, so that a cometary tail is
formed pointing away from the exciting star. A high density core
develops at the head of the globule, which could collapse to form one or
more stars. The pillars arise when an ionization front expands into a
dense region embedded in the surrounding molecular cloud, so that a cometary tail or
pillar forms due to the shadowing behind the dense clump. The pillar so
formed is supported against radial collapse by magnetic field but
undergoes longitudinal erosion by stellar winds and radiation. There
exist multiple analytical \citep{bertoldi1989,bertoldi1990} and
numerical models \citep{lefloch1994,Bisbas2011}, which explain and
reproduce many of the observed features of these structures sculpted by
the radiation and wind from massive O/B stars starting from either an
isolated or an embedded pre-existing dense clump or core.  Additionally,
there are observations which find the presence of sequential star
formation in such structures consistent with triggered formation of
stars due to the photoionization-induced shocks
\citep{sugitani2002,ikeda2008}. However, numerical simulations are yet
to provide any strong evidence supporting the enhanced formation of
stars due to triggering by the ionization fronts. Physically the
scenario is complex due to the simultaneous influence of radiation,
magnetic field and turbulence on the evolution of such clouds. Since the
pillars and globules are regions in which the gas is strongly influenced
by radiation, observation of the emission from the photon dominated
regions (PDRs) as traced by the \CII\ at 158\,\micron, \OI\ at
63\,\micron\ and mid- and high-$J$ CO lines are ideal to constrain the physical
conditions of these regions.

\begin{figure}[h]
\begin{center}
\includegraphics[width=0.5\textwidth]{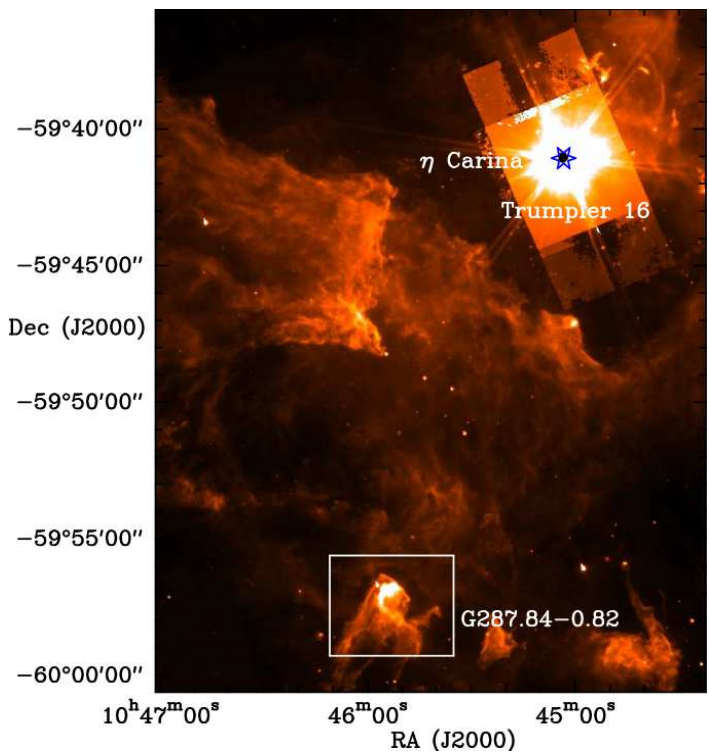}
\caption{{\it Spitzer} IRAC 8~\micron\ image of Southern pillars of Carina showing
the location of the Treasure Chest region on a larger perspective. The image artifacts from
the exceedingly bright  O2 star $\eta$\,Carina, marked by a star symbol, completely hides
the massive OB cluster, Trumpler 16, located just southwest of it. The G287.84 pillar is shown with
a rectangle.
\label{fig_carina8um}}
\end{center}
\end{figure}

The Carina Nebula Cluster (NGC\,3372) is the nearest ($d$=2.3\,kpc)
massive star-forming region in the southern hemisphere with more than 65
O stars \citep{smith2010}. Early surveys in molecular lines and
far-infrared continuum of the central parts of the Carina Nebula
suggested that the radiation and stellar winds from hot massive stars in
the region are clearing off the remains of their natal cloud
\citep{harvey1979,degraauw1981,ghosh1988}. The south eastern part of the
nebula is particularly rich in large elongated bright pillars, all of
which seem to point towards $\eta$\,Carina and Trumpler 16, a massive
open cluster with four O3-type stars, as well as numerous late O and B
stars.  Studies of this region, called the South Pillars,  in the
thermal infrared and radio suggest that it is a site of ongoing and
possibly triggered star formation \citep{smith2000, rathborne2004,
smith2005}.  Since the pillars in this region have the bright parts of
their heads pointing towards $\eta$\,Car and their extended tails
pointing away from it, it seems clear that they are sculpted by the
radiation and winds from the massive stars associated with Trumpler 16
and $\eta$\,Car.  

\begin{figure}[h]
\begin{center}
\includegraphics[width=0.5\textwidth]{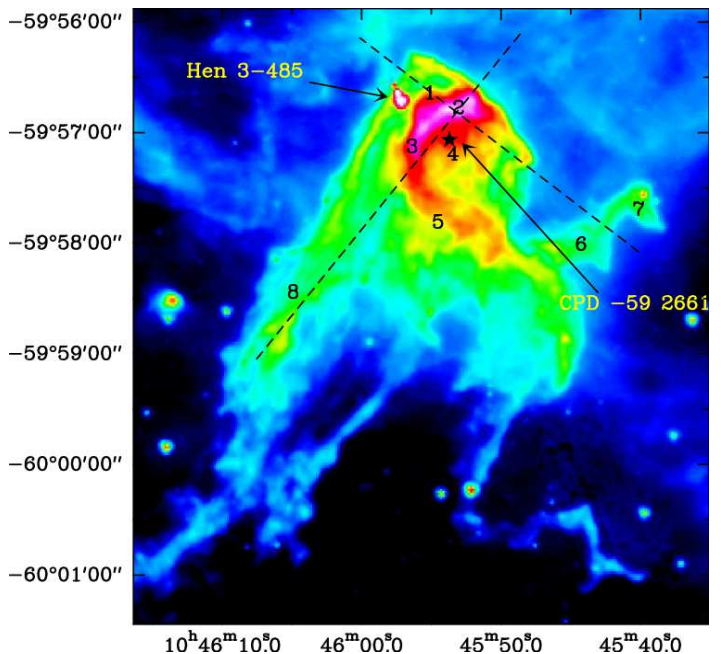}
\caption{Spitzer 8\,\micron\ image of the cometary globule
G287.84-0.82. One can see three tails extending to the south from the
head part of the globule.  All tails point away from  $\eta$\,Car and
the O stars in Trumpler 16, which are to the north west of the globule,
see Fig.~\ref{fig_carina8um}. The internal PDR, illuminated by  CPD
-59\arcdeg 2661 in the Treasure chest cluster, is seen as a bright semi
shell in the northern part of the image.  A secondary pillar, the Duck
head, is  west of the Treasure chest cluster. It also has an embedded
young star in the head. Faint foreground PDR emission is seen all
around the globule, especially to the north. We have marked the
position of  CPD -59\arcdeg 2661 and Hen 3-485. The two dashed lines
mark the cuts for position-velocity plots discussed later in the text.
The numbers 1--8 mark the positions where we extracted the spectra
discussed in Section~\ref{sec-Results}.
\label{fig_g287_8um}
}
\end{center}
\end{figure}

One of the cometary globules in the South Pillars, G287.84-0.82, is
known to have a dense compact cluster embedded in the head of the
globule. This cluster, called the Treasure Chest \citep{smith2005},
contains more than 69 young stars.  \citet{oliveira2018} estimate a
cluster age of 1.3  Myr. The most massive cluster member is CPD
-59\arcdeg 2661 (at $\alpha_{\rm J2000}$= 10h 45m 53\psec713,
$\delta_{\rm J2000}$ = -59\arcdeg 57\arcmin 3\ptsec8), an O9.5 star
\citep{walsh1984,hagele2004}, which illuminates a compact \HII\ region
seen in H$\alpha$ and in PDR  emission \citep{thackeray1950,smith2005},
see Fig.~\ref{fig_carina8um} \& \ref{fig_g287_8um}.  To the north and
east of the star the surrounding dense molecular cloud prevents the
\HII\ region from expanding. Here the distance from the expanding \HII\
region to the star  is only $\sim$ 12\arcsec, while it appears to be
expanding freely to the west, where the gas densities are low
(Fig~\ref{fig_g287_8um}).  The bright star $\sim 30$\arcsec\ northeast
of CPD -59\arcdeg 2661 (saturated in  IRAC images) is probably a member
of the embedded cluster.  This star (Hen 3-485 = Wra 15-642) is  Herbig
Be star \citep{gagne2011}. It is reddened by  $\sim$ 2 mag and has
strong infrared excess. It is associated with an H$_2$O  maser with a
single maser feature at the same velocity of the globule
\citep{breen2018}. The vibrationally excited H$_2$ observations of
Treasure Chest by \citet{hartigan2015} show the external PDR
illuminated by $\eta$\,Car.  They also show an internal PDR excited by
CPD -59\arcdeg 2661, which expands into the head of the globule. Thus,
while the cometary globule around the Treasure Chest cluster appears to
have been shaped by the radiation from the massive stars, the embedded
young stellar cluster illuminates its own PDR in the head of the
globule and will eventually dissociate the surrounding molecular cloud. 

Here we present mapping observations of the \CII\ at 158\,\micron, \OI\ at
63\,\micron, and CO emission of G287.84-0.82  in Carina to
derive a better understanding of the physical conditions of the globule and 
the Treasure Chest cluster.

\section{Observations}

\subsection{SOFIA} We have retrieved \CII\ and  CO(11--10) observations
of G287.84-0.82 (the Treasure Chest) from the data archive of the
Stratospheric Observatory for Infrared Astronomy
\citep[SOFIA;][]{young2012}. The observations (PI: X. Koenig) were part
of a larger program (01\_0064) on observing irradiated pillars in three
massive star forming regions. The observations were done with the German
REceiver for Astronomy at Terahertz frequencies
\citep[GREAT\footnote{The development of GREAT was financed by the
participating institutes, by the Federal Ministry of Economics and
Technology via the German Space Agency (DLR) under Grants 50 OK 1102, 50
OK 1103 and 50 OK 1104 and within the Collaborative Research Centre 956,
sub-projects D2 and D3, funded by the Deutsche Forschungsgemeinschaft
(DFG).};][]{heyminck2012} in the L1/L2 configuration on July 22nd 2013
during the New Zealand deployment.  These observations were made on a 70
minute leg at an altitude of 11.9 km. The L1 mixer  was tuned to
CO(11--10),  while the L2 mixer was tuned to the \CII\  $^2{\rm P}_{3/2}
\to\ ^2{\rm P}_{1/2}$ transition, see Table~\ref{tbl-1} for observation
details. The backends were fast Fourier transform spectrometers (AFFTS)
\citep{Klein12}  with 8192 spectral channels covering  a bandwidth of
1.5 GHz, providing a frequency resolution of  183.1 kHz. The Treasure
Chest was mapped in total-power-on-the-fly mode with a 1 second
integration time and sampled every 8\arcsec\ for a map size of
248\arcsec\ $\times$ 168\arcsec.  The map was centered on ($\alpha$,
$\delta$) = (10$^{\rm h}$ 45$^{\rm m}$ 55\fs05, -59\degr\
57\arcmin41\farcs2) (J2000) with an off-position 5\arcmin\ to the south.
All offsets mentioned in the text and figure captions hereafter
refer to this centre position. Here we only use the CO(11--10) data,
because the \CII\ observations made with upGREAT, see below, are of
superior quality. 

The source G287.84-0.82  was also observed with  upGREAT in consortium time from
Christchurch on 2018 June 14 on a 65 minute leg at an altitude from
11.4--12.4 km. The upGREAT was in the Low Frequency Array/High
Frequency Array (LFA/HFA) configuration with both arrays operating in
parallel.  The 14 beam LFA array \citep{Risacher16} was tuned to \CII,
while the 7 beam HFA array was tuned to the $^3$P$_{1}$--$^3$P$_{2}$
63\,\micron\  \OI\ line. The spectrometer backends are the last
generation of fast Fourier transform spectrometers (FFTS;
\citet{Klein12}). They cover an instantaneous intermediate frequency
(IF) bandwidth of 4 GHz per pixel, with a spectral resolution of 142
kHz. To ensure good baseline stability for the \OI\ line we mapped the
Treasure Chest in Single Beam Switch  (SBS) on-the-fly mode by chopping
to the east of the cometary globule. The region we chopped to appears
free of \CII\ emission, even though there is widespread faint
foreground emission at velocities of $\sim$ -30 to -24 \kms\ in the
area around the globule. We split the region into two sub-maps. The
northern part of the globule was done as a 123\arcsec\ $\times$
69\arcsec\ map with a chop amplitude of 120\arcsec\ at a position angle
(p.a.) of 60\degr. The map was centered with an offset of
0\arcsec,+27\arcsec\ relative to our track position and sampled every
3\arcsec\ with an integration time of 0.4 seconds.  The second map had
a size of 231\arcsec\ $\times$ 93\arcsec and was centered at an offset
of -9\arcsec,-54\arcsec. For this map we used a chop amplitude of
135\arcsec\ at a p.a. of 84\degr. This setup provided us a fully
sampled map in \OI\ over most of the globule and an oversampled map of
\CII\ with excellent quality. 

The atmospheric conditions were quite dry. The zenith optical depth for
\CII\ varied from $\sim$0.19 at the beginning of the leg to $\sim$0.15
at the end of the leg. The system temperatures for most pixels were
about 2400 K, some around 2900 K, and three outliers had higher system
temperatures, 3300 - 4000 K. One noisy pixel was removed in the post
processing. Beam efficiencies were applied individually to each pixel,
see Table~\ref{tbl-1}.  For  \OI\ the zenith optical depth varied
between 0.24 to 0.43 and system temperatures ranged from 2900 to 4650\,K
excluding one noisy pixel, which was not used.

All the GREAT data were reduced and calibrated by the GREAT team, including
correction for atmospheric extinction \citep{Guan12} and calibrated in antenna
temperature scale,T$_{\rm A}^*$, corrected for a forward scattering efficiency
of 0.97, and in main beam antenna temperature, T$_{\rm mb}$.

The final data cubes were created using CLASS\footnote{CLASS is part of
the Grenoble Image and Line Data Analysis Software (GILDAS), which is
provided and actively developed by IRAM, and is available at
http://www.iram.fr/IRAMFR/GILDAS}. Maps were created  with a velocity
resolution of 0.5~\kms. The rms noise per map resolution
element with 0.5 \kms\ velocity resolution is $\sim$ 1.3 K
for the  CO(11--10) map,  0.70 K for \CII,  and  2.4 K for \OI.

\subsection{APEX}

G287.84-0.82 was observed  on March 28, 2017 using the PI230 receiver
on the 12~m Atacama Pathfinder EXperiment (APEX\footnote{APEX, the
Atacama Pathfinder Experiment is a collaboration between the
Max-Planck-Institut f\"ur Radioastronomie at 55\%, Onsala Space
Observatory (OSO) at 13\%, and the European Southern Observatory (ESO)
at 32\% to construct and operate a modified ALMA prototype antenna as a
single dish on the high altitude site of Llano Chajnantor. The
telescope was manufactured by VERTEX Antennentechnik in Duisburg,
Germany}) telescope, located at Llano de Chajnantor in the Atacama
desert of Chile \citep{Gusten06}. PI230 is a dual sideband separating
dual polarization receiver. Each band has a bandwidth of 8 GHz and
therefore each polarization covers 16 GHz. Each band is connected to
two Fast Fourier Transform fourth Generation spectrometers (FFTS4G)
\citep{Klein12} with 4  GHz bandwidth and 65536 channels, providing a
frequency resolution of 61 kHz, i.e., a velocity resolution of $\sim$
0.079 \kms\ at 230 GHz. The PI230 receiver can be set up to
simultaneously observe CO(2--1), $^{13}$CO(2--1), and C$^{18}$O(2--1).
This frequency setting also includes three H$_2$CO lines and one SO
line in the lower sideband, see Table~\ref{tbl-1}, all of which were
detected in G287.84-0.82.

G287.84-0.82 was mapped in on-the-fly total power mode and sampled every
10\arcsec\ with an integration time of 0.5 seconds. The map size was
$\pm$100\arcsec\ in RA and from -150\arcsec\ to +95\arcsec\ in
Dec.  The off position used in these observations was  at RA =10$^{\rm
h}$ 49$^{\rm m}$ 14$^{\rm s}$, Dec = -59\degr\ 35\arcmin40\arcsec{}
(J2000).  The weather conditions were quite good for 1.3 mm
observations. The zenith optical depth varied from 0.146 - 0.156 at 230
GHz resulting in a system temperature of $\sim$ 155 K.

The final data cubes were created using CLASS by resampling the spectra
to 0.5 \kms\ velocity resolution.The rms noise per map resolution element
(half a beam width) is $\sim$ 0.18 K.

\begin{table}[h]
\begin{center}
\caption{Observation Details: Frequencies, Receivers, Beam sizes, and beam efficiencies\label{tbl-1}} 
{\scriptsize
\begin{tabular}{llllcl}
\hline\hline
Telescope & Receiver & Transition &  Frequency   & $\theta_{HPBW}$ & $\eta_{mb}$ \\
&  & & [GHz] & [~\arcsec{}~]  &  \\
\hline
SOFIA & GREAT L1 & CO(11--10) &  1267014.49 & 21.0 & 0.67\\
& LFA & \CII\ $^{2}P_{3/2}\rightarrow^{2}P_{1/2}$	 & 1900536.90 & 14.0 & 0.67$^a$\\
& LFA & \thCII\ ($F$=2--1) & 1900466.10 & 14.0 & 0.67$^a$\\
& HFA & \OI\ $^{3}P_{0}\rightarrow^{3}P_{1}$& 4744777.49  & 6.3 & 0.65$^a$\ \\
APEX & PI230 & CO(2--1) & 230538.000 & 27.3& 0.68 \\
&&  \thCO(2--1) & 220398.684 & 28.5& 0.68 \\
&&  \CeiO(2--1) & 219560.354 & 28.6& 0.68 \\
&&  H$_2$CO(3$_{03}$--2$_{02}$) & 218222.192 & 28.8 & 0.68 \\
&&  H$_2$CO(3$_{22}$--2$_{21}$)& 218475.630 & 28.8 & 0.68 \\
&&  H$_2$CO(3$_{21}$--2$_{20}$)& 218760.066 & 28.8 & 0.68 \\
&& SO (6$_5$--5$_4$) & 219949.44 & 28.6 & 0.68 \\
\hline
\hline
\end{tabular}}
\end{center}
{\noindent $^a$ median across array pixels}

\end{table}

\section{Results}
\label{sec-Results}

\begin{figure*}[h]
\begin{center}
\includegraphics[width=0.8\textwidth]{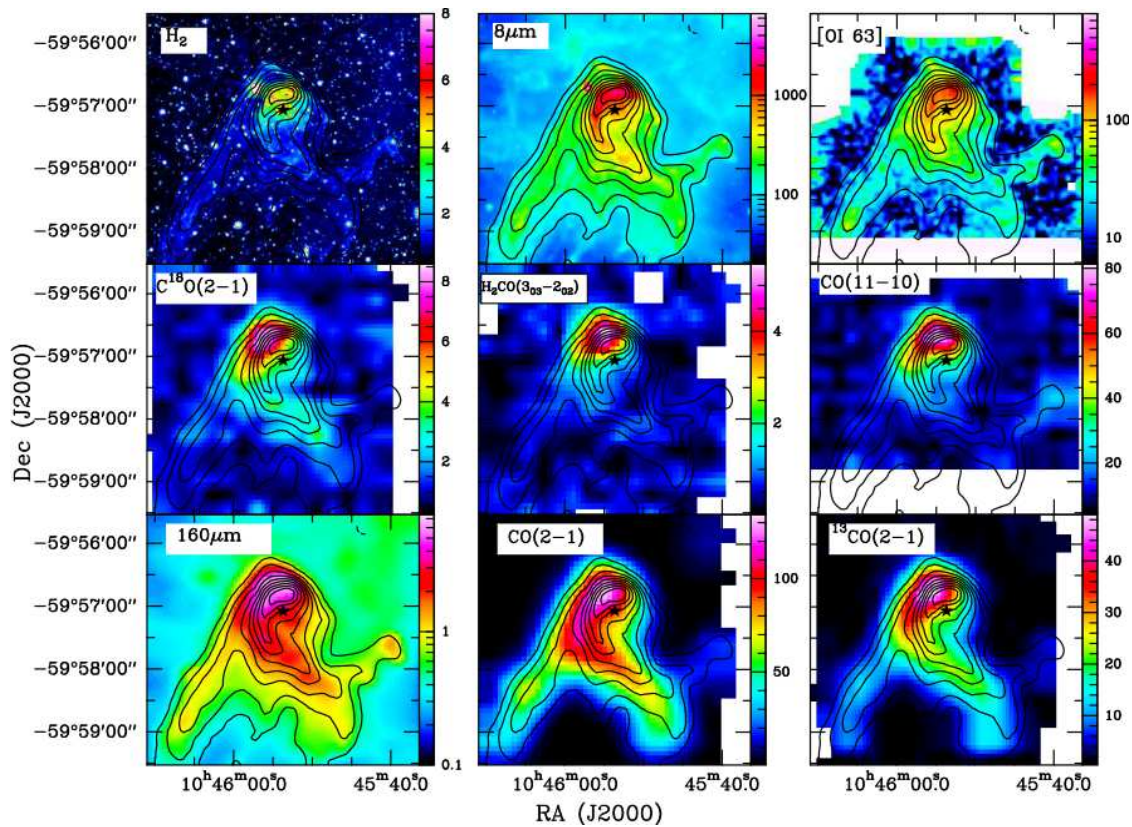}
\caption{Contours of intensity integrated between -20 and -5\,\kms\ of \CII\ 
(black contour) overlayed on false-color images of tracers mentioned in
the panels. The top row corresponds to tracers of PDR gas, the middle row
to tracers of high density gas and the bottom row to the tracers of
overall gas and dust distribution in the region. The filled asterisk shows
the position of CPD-59\arcdeg 2661.  The color scale for each panel
is shown next to the panel and the units are K\,\kms\ for the spectral
lines, MJy\,sr$^{-1}$ for 8\,\micron\ and Jy/pixel for PACS (3\arcsec\
pixel).  The values of \CII\ contours are 10 to 100\% of the peak 
intensity of 505\,K~\kms. 
\label{fig_overlays}}
\end{center}
\end{figure*}

Figure~\ref{fig_overlays} shows a comparison of the integrated intensity image
(in K~km$^{-1}$) of  \CII\ at 158 \micron\ (contours) overlaid on  fluorescent
H$_2$, 8 $\mu$m PAH emission, 160\,\micron\ dust continuum as well as on
integrated intensity images of most spectral lines that we have mapped. The CO,
\thCO\ and 160\,\micron\ dust continuum  images show that  the G287.84-0.82
cloud has the shape of a cometary globule with three spatially separated
extensions/tails in the south, which all point away from  $\eta$\,Car and
Trumpler 16.  The $\eta$\,Car is to the northwest, approximately at a projected
distance of 12 pc from G287.84-0.82.  There is a secondary cloud surface
$\sim$60\arcsec\ south of CPD -59\arcdeg 2661, which is also roughly
perpendicular to $\eta$\,Car. On the western side there is a narrow second
pillar with a head that looks like the head of a duck, see Fig. 3 in
\citet{smith2005} and Fig.~\ref{fig_g287_8um}. The structure of the cloud is
consistent with it being exposed to and shaped by the UV radiation and stellar
winds from $\eta$\,Car and from the O3 stars in Trumpler  16.  The \CII\
emission matches extremely well with the other bona fide PDR tracers like the
fluorescent H$_2$, 8\,\micron\ PAH and \OI\ 63\,\micron\ emission, suggesting
that most of the \CII\ emission originates in the irradiated cloud surfaces. The
PDR gas inside the pillar forms a shell around the Treasure Chest cluster. Here
the emission appears to be dominated by the stellar cluster with only a minor
contribution from the externally illuminated cloud surfaces.  High density
tracers like CO(11--10), C$^{18}$O(2--1) and \formaldehydethtwo\ and SO (the
latter only shown as a velocity-channel map in Fig.\,\ref{fig_sochan}) all peak
to the north-east of the positions of both the PDR peak and the embedded star
cluster.  The distribution of intensities and the relative location of the high
density peak suggest that the head of the cloud is more compressed towards the
north-eastern edge and the PDR shell curving around the Treasure Chest cluster
shows that the expanding \HII\ region both heats up and compresses the
surrounding cloud, especially to the north east. Clear detection of 
\formaldehyde\ and CO(11--10) at the position of the column density peak suggest
gas densities in excess of 10$^5$\,\cmcub.

\begin{figure}[h]
\begin{center}
\includegraphics[width=0.5\textwidth]{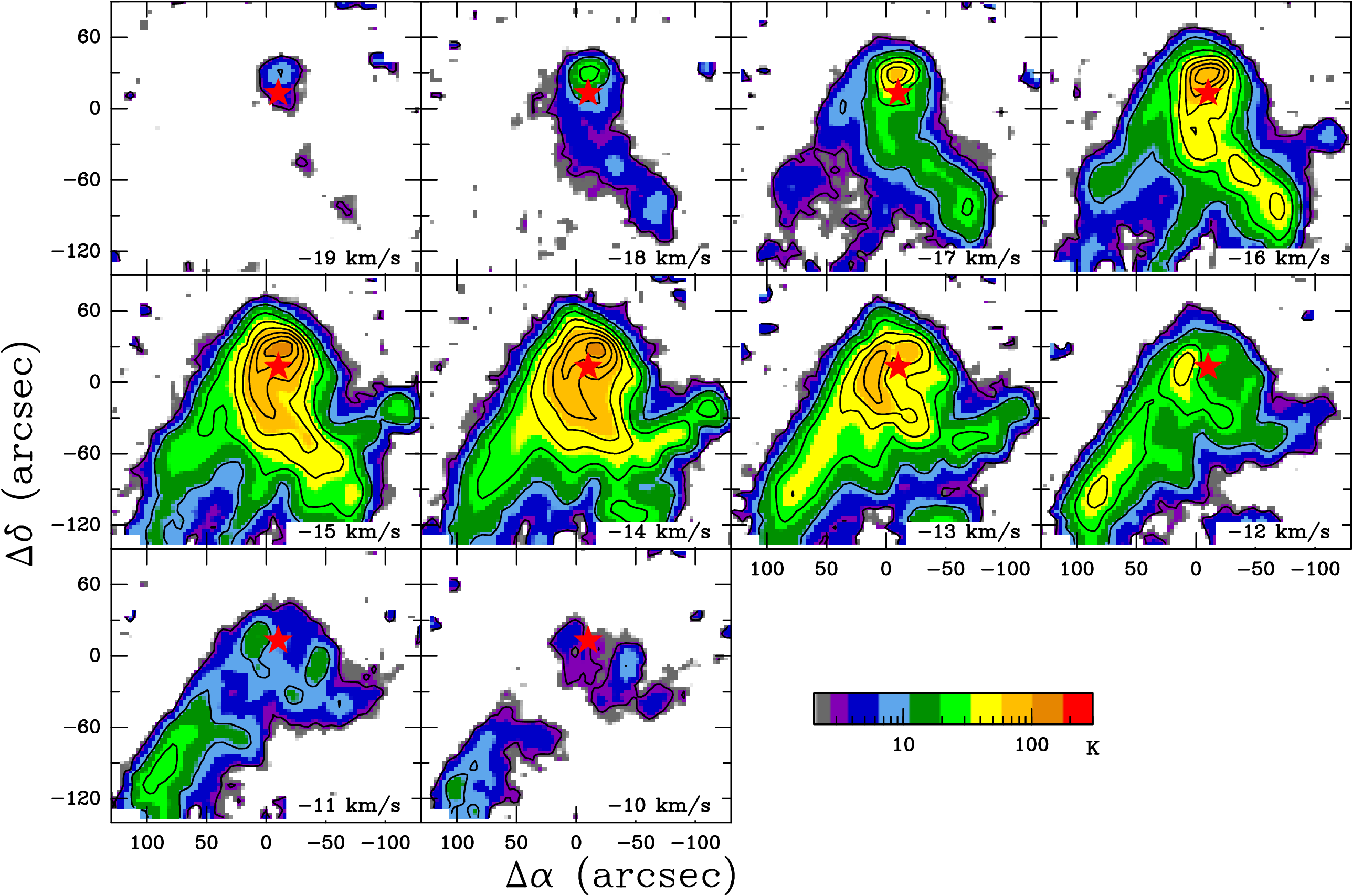}
\includegraphics[width=0.5\textwidth]{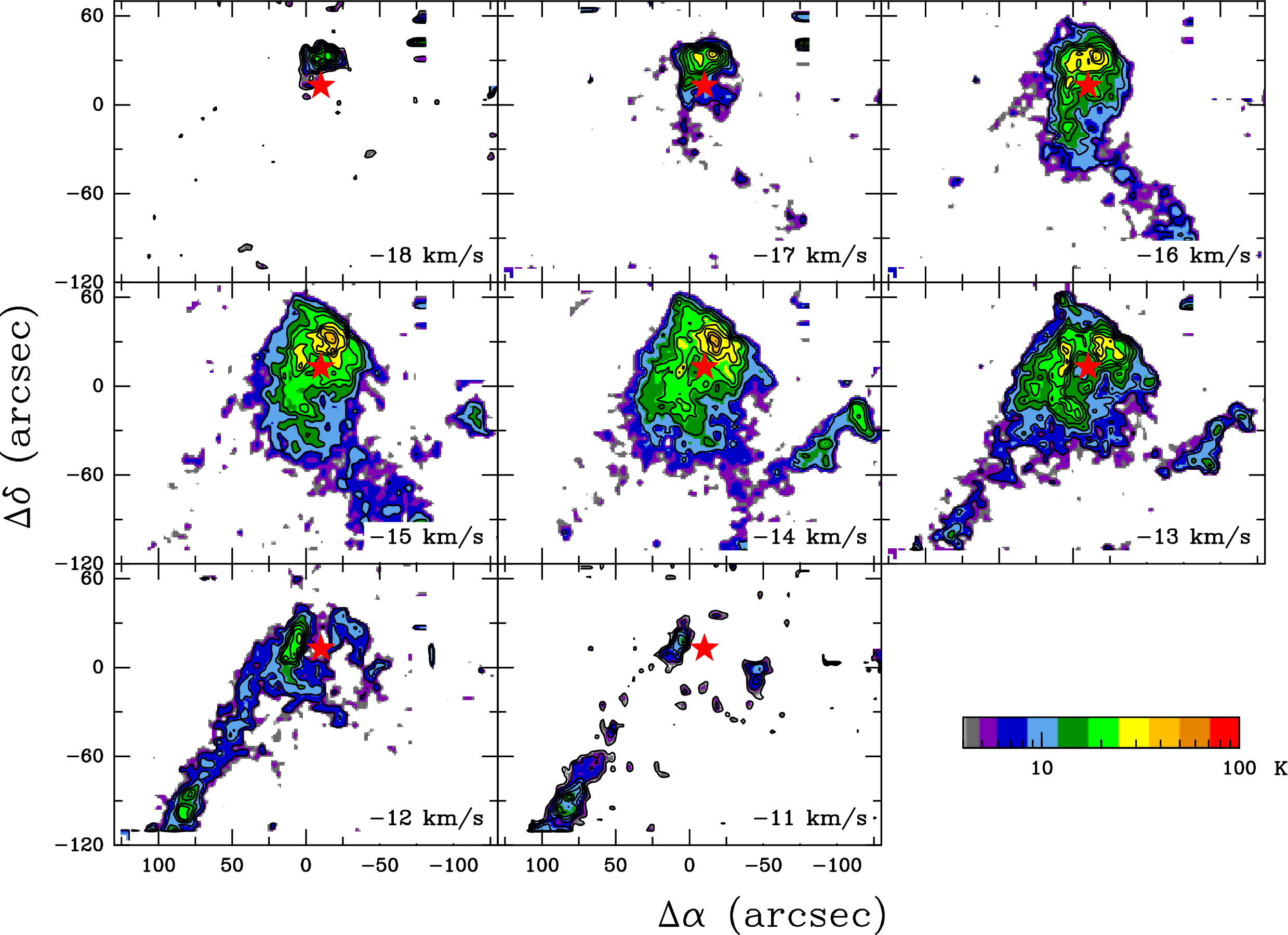}
\caption{Channel maps of \CII\ (top) \& \OI\ 63\micron\ (bottom)
emission. Velocities corresponding to the channel are marked in each
panel. The red star marks the position of CPD-59\arcdeg2661. The
positional offsets are relative to the center
$\alpha$=10$^h$45$^m$55\fs05,$\delta$=-59$^d$57$^m$16\farcs7 (J2000).
\label{fig_chanmaps}}
\end{center}
\end{figure}

Figure\,\ref{fig_chanmaps} shows the channel maps for the \CII\ and \OI\
63\,\micron\ emission. Both maps show that the bulk of the gas in the head of
the region is at a velocity of $\sim$-14.5 \kms, while the tails are generally
red-shifted.  This suggests that the globule is somewhat behind $\eta$\,Car,
which has a radial velocity of -20 \kms\ \citep{smith2004}, so that the stellar
wind and the radiation pressure from the star push them in the radial direction
from $\eta$\,Car, making them appear red-shifted. The higher angular resolution
of the \OI\ 63\,\micron\ observations (particularly around $\upsilon=$-13\,\kms
in Fig.\,\ref{fig_chanmaps}) helps identify the dense PDR shell around CPD
-59\arcdeg 2661. The \CII\ channel maps also show a tail-like structure to the
east at a velocity of $\sim$ -15~\kms, which is not seen in \OI\ 63\,\micron\
or CO(11--10) emission (Fig.\,\ref{fig_co11chan}).  The comparison with channel
maps of all observed CO, H$_2$CO and SO lines (Figs\,\ref{fig_co11chan},
\ref{fig_co21chan}, \ref{fig_13cochan}, \ref{fig_c18ochan}, \ref{fig_h2cochan},
and \ref{fig_sochan}), show that the "eastern tail-like" -14.8 \kms\ component
is clearly detected  in the CO(2--1) channel maps and only partially in the
\thCO(2--1) channel maps.  This suggests that the -14.8 \kms\ emission component
in the tail has lower density, since it is not detected in the transitions,
which either (a) have high critical density like \OI, CO(11--10) and
\formaldehyde\ or (b) trace regions of high column density e.g., \CeiO(2--1). 
This gas is probably part of the halo around the globule, which has not yet been
affected by the wind from $\eta$\,Car.

\begin{figure}[h]
\begin{center}
\includegraphics[width=0.5\textwidth]{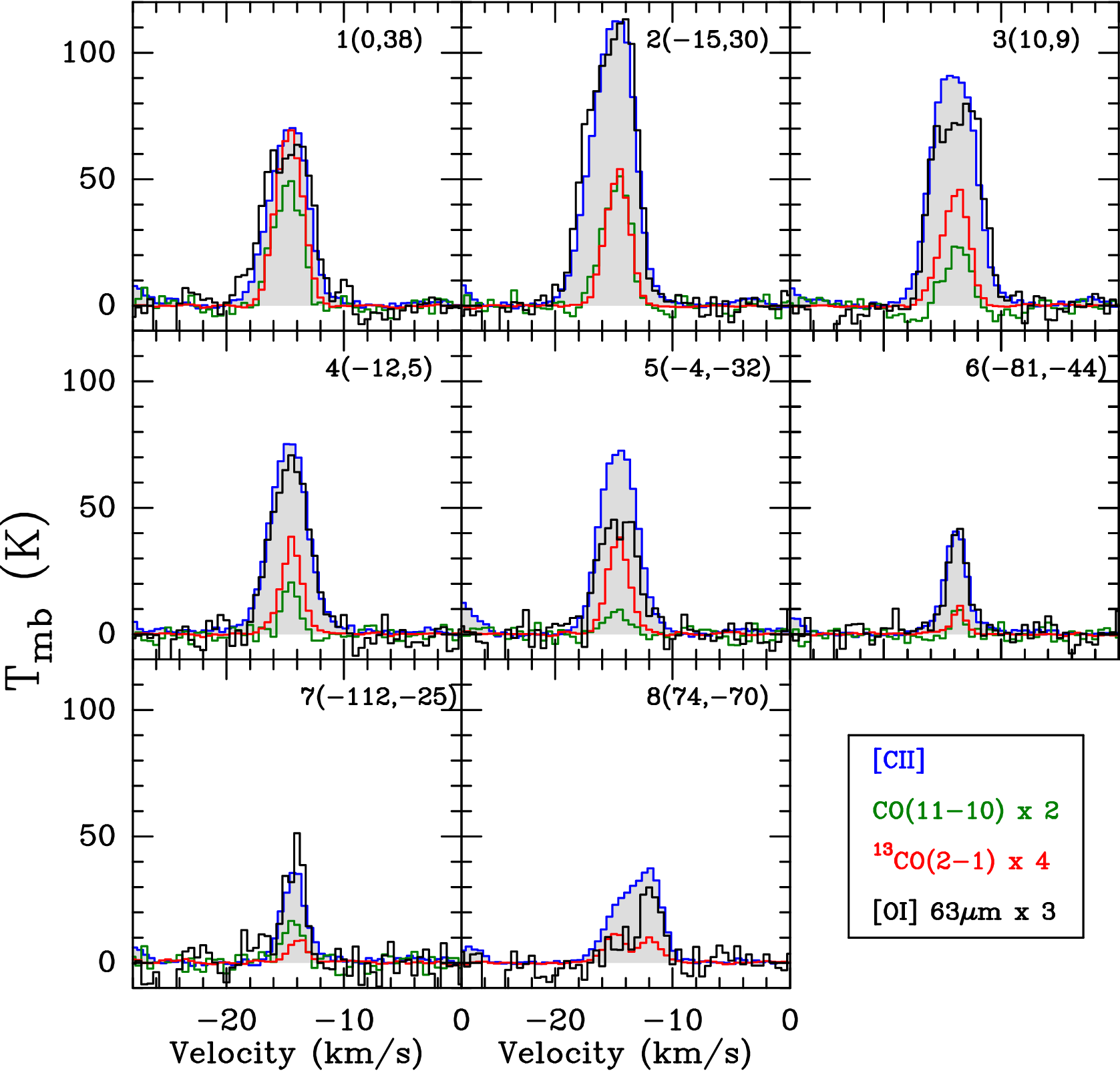}
\caption{Comparison of emission spectra of \CII\ (filled histogram),
\thCO(2--1)(red), CO(11--10) (green) and \OI\ 63\,\micron\ (black) at
selected positions shown in panel 2 of Fig.\,\ref{fig_overlays}.  The
\thCO, CO(11--10), and \OI\ spectra are multiplied by factors of  4, 2 and
3, respectively.
\label{fig_selspec}}
\end{center}
\end{figure}

\begin{table}[h]
\caption{Results of single (or double for position \#8) component Gaussian 
fitting of observed spectra \label{tab_gaussfit}}
{\small
\begin{tabular}{clrrr}
\hline
($\Delta \alpha$,  $\Delta \delta$) & Transition &  $I$\phn\phn\phn\phn &
$\upsilon_{\rm LSR}$\phn\phn\phn & $\Delta \upsilon$\phn\phn\phn \\
& & K\,\kms & \kms\phn\phn & \kms\phn\\
\hline
(0\arcsec,38\arcsec) & \CII &  288.4$\pm$6.2 & -14.7$\pm$0.1 & 3.8$\pm$0.1\\
          & CO(2--1)       &  137.4$\pm$0.3 & -14.8$\pm$0.1 & 3.4$\pm$0.1\\
          & $^{13}$CO(2--1)&   49.8$\pm$0.2 & -14.7$\pm$0.1 & 2.7$\pm$0.1\\
          & C$^{18}$O(2--1)&    7.7$\pm$0.1 & -14.8$\pm$0.1 & 2.3$\pm$0.1\\
          & CO(11--10)     &   86.9$\pm$3.3 & -14.8$\pm$0.1 & 2.9$\pm$0.1\\
	  & \OI\ 63\micron\ & 125.7$\pm$2.7 & -15.0$\pm$0.1 & 4.9$\pm$0.1\\
\hline
 (-15\arcsec,30\arcsec) & \CII  & 511.0$\pm$6.0 & -15.0$\pm$0.1 &  4.2$\pm$0.1\\
          & CO(2--1)        & 127.7$\pm$0.3 & -14.8$\pm$0.1 &  3.2$\pm$0.1\\
          & $^{13}$CO(2--1) &  33.9$\pm$0.1 & -14.7$\pm$0.1 &  2.4$\pm$0.1\\
          & C$^{18}$O(2--1) &   4.2$\pm$0.1 & -14.7$\pm$0.1 &  2.1$\pm$0.1\\
          & CO(11--10)      &  64.0$\pm$3.9 & -14.5$\pm$0.1 &  2.5$\pm$0.2\\
	  &\OI\ 63\micron\  & 210.0$\pm$3.8 & -15.0$\pm$0.1 &  4.4$\pm$0.1\\
\hline
  (10\arcsec,9\arcsec) & \CII & 435.5$\pm$6.0 & -14.1$\pm$0.1 &   4.2$\pm$0.1\\
         & CO(2--1)        & 107.6$\pm$0.3 & -14.2$\pm$0.1 &   3.5$\pm$0.1\\
         & $^{13}$CO(2--1) &  32.8$\pm$0.2 & -13.9$\pm$0.1 &   2.7$\pm$0.1\\
         & C$^{18}$O(2--1) &   4.3$\pm$0.1 & -14.0$\pm$0.1 &   2.9$\pm$0.1\\
         & CO(11--10)      &  31.4$\pm$3.5 & -13.7$\pm$0.2 &   2.6$\pm$0.3\\
	 &\OI\ 63\micron\  & 103.3$\pm$3.9 & -13.5$\pm$0.1 &   4.3$\pm$0.2\\
\hline
 (-12\arcsec,5\arcsec) & \CII  & 303.1$\pm$4.8 & -14.7$\pm$0.1 &   3.8$\pm$0.1\\
         & CO(2--1)        &  88.8$\pm$0.3 & -14.6$\pm$0.1 &   2.9$\pm$0.1\\
         & $^{13}$CO(2--1) &  20.2$\pm$0.1 & -14.5$\pm$0.1 &   2.1$\pm$0.1\\
         & C$^{18}$O(2--1) &   2.7$\pm$0.1 & -14.4$\pm$0.1 &   2.0$\pm$0.1\\
         & CO(11--10)      &  28.4$\pm$2.8 & -14.5$\pm$0.1 &   2.4$\pm$0.3\\
	 &\OI\ 63\micron\  &  90.4$\pm$3.9 & -14.7$\pm$0.1 &  3.8$\pm$0.2\\
\hline
 (-4\arcsec,-32\arcsec) & \CII & 293.4$\pm$8.9 & -14.6$\pm$0.1 &   3.7$\pm$0.1\\
          & CO(2--1)        &  96.0$\pm$0.5 & -14.6$\pm$0.1 &   2.7$\pm$0.1\\
          & $^{13}$CO(2--1) &  22.4$\pm$0.2 & -14.6$\pm$0.1 &   2.2$\pm$0.1\\
          & C$^{18}$O(2--1) &   1.9$\pm$0.1 & -14.8$\pm$0.1 &   1.8$\pm$0.1\\
          & CO(11--10)      &   7.0$\pm$2.1 & -14.5$\pm$0.1 &   0.9$\pm$0.3\\
	  &\OI\ 63\micron\ &  67.9$\pm$4.0 & -15.0$\pm$0.1 &   3.9$\pm$0.3\\
\hline
 (-81\arcsec,-44\arcsec) & \CII  & 95.9$\pm$3.9 & -13.9$\pm$0.1 &   2.3$\pm$0.1\\
           & CO(2--1)        & 15.1$\pm$0.3  & -13.8$\pm$0.1 &   1.6$\pm$0.1\\
           & $^{13}$CO(2--1) &  3.8$\pm$0.1  & -13.6$\pm$0.1 &   1.3$\pm$0.1\\
           & C$^{18}$O(2--1) &  0.5$\pm$0.1  & -13.5$\pm$0.2 &   2.0$\pm$0.5\\
           & CO(11--10)      &  7.0$\pm$1.4  & -13.7$\pm$0.1 &   1.0$\pm$0.2\\
	   &\OI\ 63\micron\ & 19.5$\pm$3.5 & -13.7$\pm$0.2 &   1.9$\pm$0.5\\
\hline
 (-112\arcsec,-25\arcsec) & \CII  & 92.3$\pm$3.9 & -14.2$\pm$0.1 &   2.6$\pm$0.1\\
           & CO(2--1)        & 19.4$\pm$0.5  & -14.3$\pm$0.1 &   2.3$\pm$0.1\\
           & $^{13}$CO(2--1) &  4.6$\pm$0.2  & -13.9$\pm$0.1 &   1.8$\pm$0.1\\
           & CO(11--10)      & 17.1$\pm$2.8  & -14.4$\pm$0.1 &   1.7$\pm$0.3\\
	   &\OI\ 63\micron\ & 31.2$\pm$5.5 & -12.3$\pm$0.3 &   3.9$\pm$0.7\\
\hline
 (74\arcsec,-70\arcsec) & \CII  & 70.1$\pm$2.3 & -14.3$\pm$0.1 &  2.8$\pm$0.1\\
                        &       & 85.2$\pm$2.2 & -11.9$\pm$0.1 &   2.3$\pm$0.1\\
           & CO(2--1)        & 31.9$\pm$1.0  & -15.0$\pm$0.1 &   2.7$\pm$0.1\\
           &                & 33.3$\pm$0.9  & -12.0$\pm$0.1 &   2.5$\pm$0.1\\
           & $^{13}$CO(2--1) &  3.7$\pm$0.3  & -14.9$\pm$0.1 &   1.8$\pm$0.1\\
           &                &  6.8$\pm$0.3  & -12.0$\pm$0.1 &   2.3$\pm$0.1\\
	   &\OI\ 63\micron\ & 7.6$\pm$2.7 & -14.7$\pm$0.5 &   2.4$\pm$1.0\\
	   &                & 21.0$\pm$2.6 & -11.8$\pm$0.1 &   1.9$\pm$0.2\\

\hline
\hline
\end{tabular}
}
\end{table}


\begin{figure}[h]
\begin{center}
\includegraphics[width=0.4\textwidth]{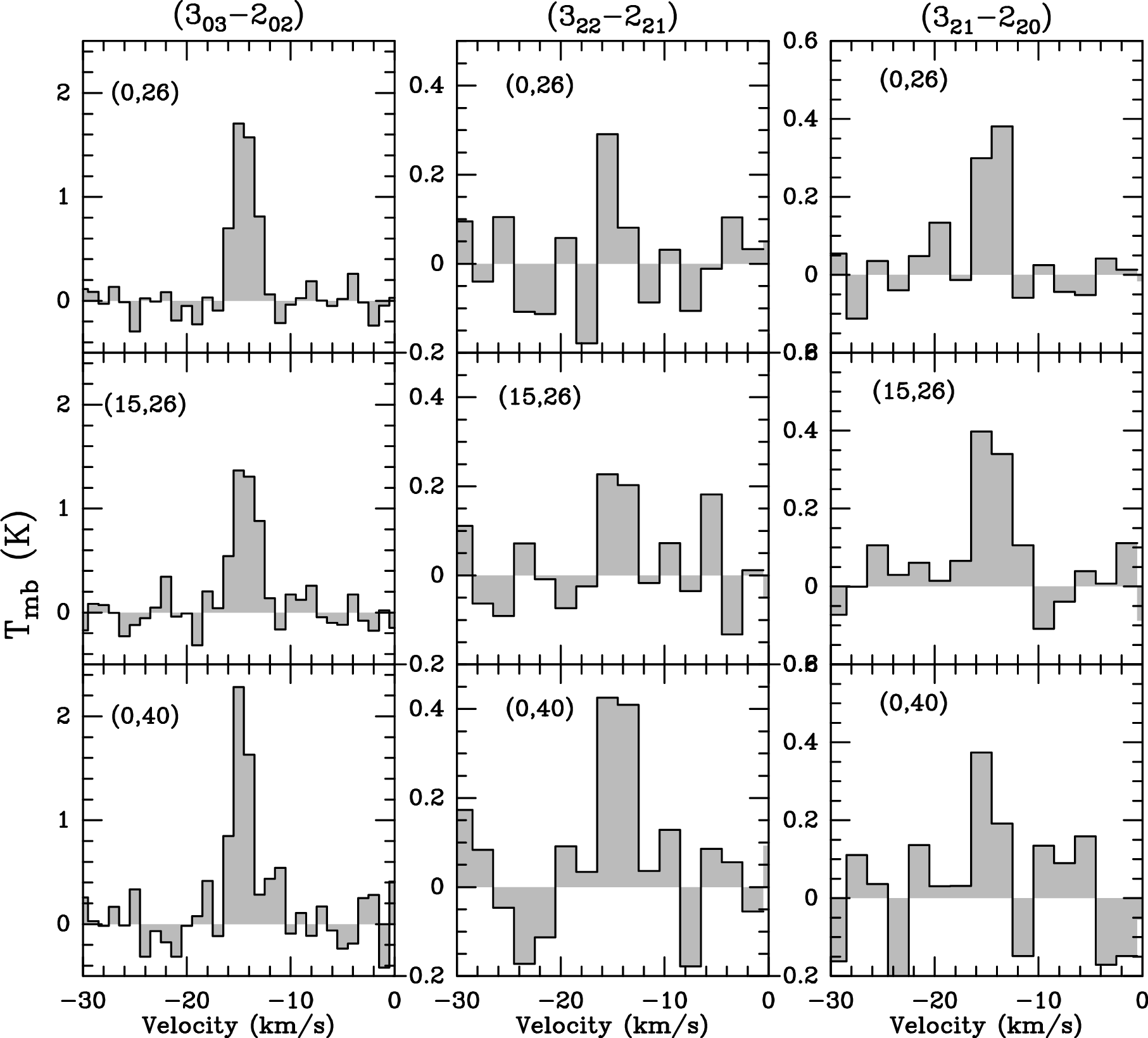}
\caption{Spectra of the $J$=(3$_{03}$--2$_{02}$), (3$_{22}$--2$_{21}$)
and (3$_{21}$--2$_{20}$) transitions of para-H$_2$CO at selected positions, 
where the fainter lines are clearly detected.
\label{fig_h2cospec}}
\end{center}
\end{figure}

We next compare the spectra of the different tracers at eight selected
positions which are marked in Fig.\,\ref{fig_g287_8um}. These positions
are selected to sample positions in the head of the globule  (\#3 to
\#5), as well as some specific features,  such as the CO peak (\#1), the
\CII\ peak (\#2), the neck (\#6), the duck head (\#7),  and the eastern
tail (\#8). Since the line profiles of the $J$=2--1 transitions of CO
(and isotopologues) are similar,  we only show the \thCO\ spectra along
with other spectra in Fig.\,\ref{fig_selspec}.  At the position of the
column density peak (\#1), we find that all spectra have similar line
widths, with \OI\ being slightly broader than the others.  At position
\# 2 the PDR tracers \CII\ and \OI\ are significantly broader than the
molecular gas tracers.  Table\,\ref{tab_gaussfit} gives the result of
fitting the spectra using a single Gaussian velocity component for all
positions except position \#8 where we have fitted two Gaussian velocity
components. We note that for positions \#7 and \#8 only the tracers that
are detected are presented in Table\,\ref{tab_gaussfit}.

We find that except for position \#1, the \CII\ and \OI\ lines are
broader than the CO lines by 1--1.5~\kms.  The  line widths and
spectral line profiles of \CII\ and \OI\ match very well, suggesting
that both \OI\ and \CII\ emission originate in PDR gas with no
significant contribution from (a) shocks to \OI\ and (b) the \HII\
region to the \CII\ emission. A closer inspection of the \OI\ and \CII\
data in the head of the globule suggests that although the outer
profiles of the spectra from the two species match well, the \OI\
spectra show a flattened or double-peaked profile close to the column
density peak and further south-east (e.g. positions 1, 3 and 5 in
Fig.\ref{fig_selspec}), whereas the \CII\ spectra always show a smooth
single-peaked Gaussian profile. The two velocity components detected in
CO, \thCO, \CII, and \OI\ at position \#8, correspond to the main
globule and the red-shifted eastern tail. 

Figure~\ref{fig_h2cospec} shows para-H$_2$CO spectra at three positions
where all the three transitions  $J$=(3$_{03}$--2$_{02}$),
(3$_{22}$--2$_{21}$) and (3$_{21}$--2$_{20}$) are detected with good SNR. The
strongest transition, $J$=(3$_{03}$--2$_{02}$), is detected over a much
larger area around the column density peak (as shown in
Fig.\,\ref{fig_overlays} and \ref{fig_sochan}). At the positions (0,26)
and (15,26), the para-\formaldehyde\  3$_{03}$--2$_{02}$ transition shows
a single peaked profile, while at (0,40), where the line is the
strongest, there is a hint of a red-shifted component.  The position of
the peaks of the two fainter transitions match with the brighter peak. The
widths of the \formaldehyde\ lines match well with the line widths of the
other molecular lines at these positions.  The emission from the
(6$_5\rightarrow5_4$) transition of SO spans an even narrower range of
velocities and is mostly confined to the higher density head region of the
globule (Fig.~\ref{fig_sochan}).

\begin{figure}[h]
\begin{center}
\includegraphics[width=0.4\textwidth]{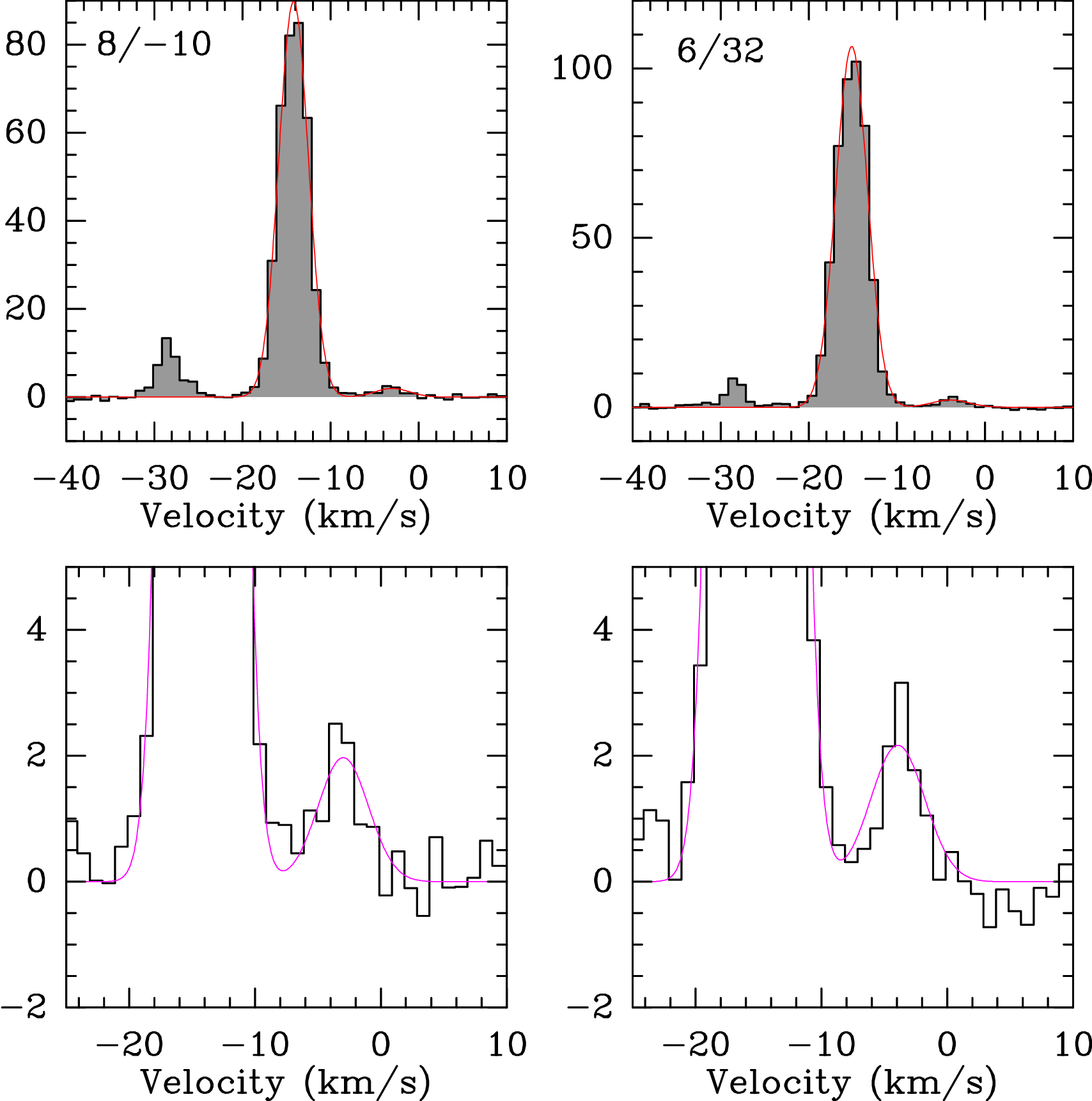}
\caption{Examples of \CII\ spectra at two positions where \thCII\ is also
detected. The emission feature at $\sim$ -28 \kms\ is from faint foreground PDR
emission unrelated to the globule. The lower panels show a zoomed in view of the
\thCII\ lines, along with the result of simultaneous fitting of \CII\ and
\thCII\ lines as described in the text.
\label{fig_13cplus}}
\end{center}
\end{figure}

The high S/N of the observed \CII\ spectra allows us to simultaneously
detect the brightest hyperfine component, F = 2--1, of the
$^{13}$\cplus\ fine structure line at 50 positions in the area we have
mapped. If we average these spectra together, we also see the two
fainter hyperfine transitions F = 1--0 and F = 1--1 of \thCII.
However, they are too faint in individual positions for a meaningful
analysis, and we therefore only use the F = 2--1 transition for this
purpose.  Fig.~\ref{fig_13cplus} shows examples  of \thCII\ F = 2--1
line, which is offset by +11.2  \kms\ from the \CII\ line, at two
selected positions. The line-integrated intensity of the \thCII\ line
was obtained by fitting the \CII\ and \thCII\ lines simultaneously.  In
this fitting procedure the brighter \CII\ line is fitted first and the
fainter \thCII\ line is constrained to be located at a velocity
relative to the brighter component given by the velocity separation of
the two transitions.

\section{Analysis}

\subsection{LTE \& Non-LTE analysis of molecular emission
\label{sec_tkinmol}}

We derive a first order estimate of the total gas column density using
the observed integrated line  intensities of CO and $^{13}$CO $J$=2--1.
Since the CO(2--1) emission is mostly optically thick, we use the
maximum temperatures for CO(2--1) as the kinetic temperature ($T_{\rm
kin}$). We further assume that $T_{\rm ex}$ for all the isotopes of CO
are identical and equal to $T_{\rm kin}$. For all pixels where both CO
and $^{13}$CO is detected we estimate the $N$(\thCO) to lie between
(1.3--76)$\times 10^{15}$\,\cmsq. Using a $^{12}$CO/$^{13}$CO ratio of
65 \citep{rathborne2004}, and CO/H$_2$ = 10$^{-4}$, we find $N$(H$_2$)
to range between 8.5$\times 10^{20}$ -- 4.9$\times 10^{22}$\,\cmsq.
This estimate of $N$(H$_2$) agrees well with what \citet{breen2018}
found from observations of NH$_3$ (1,1) and (2,2) with the Australia
Compact Array (ATCA), 2.2$\times$10$^{22}$ $\pm$
0.9$\times$10$^{22}$~\cmsq, as well as with the estimate by
\citet{rebolledo2016} based on observations of $J$=1--0 transitions of
$^{12}$CO and $^{13}$CO. It also matches reasonably well with the
column density estimated by both \citet{roccatagliata2013} and
\citet{schneider2015} based on dust continuum observations with the
{\it Herschel} Space Observatory.  Our estimate of the CO column
density, 8.5$\times 10^{16}$ -- 4.9$\times 10^{18}$\,\cmsq, is also
consistent with the results by \citet{rathborne2004}, who used CO(2--1)
\& CO(1--0) observed with SEST to derive $T_{\rm ex}$ = 40 K and
$N$(CO) = 1.4$\times 10^{18}$\,\cmsq.

%

\begin{figure}[h]
\begin{center}
\includegraphics[width=0.5\textwidth]{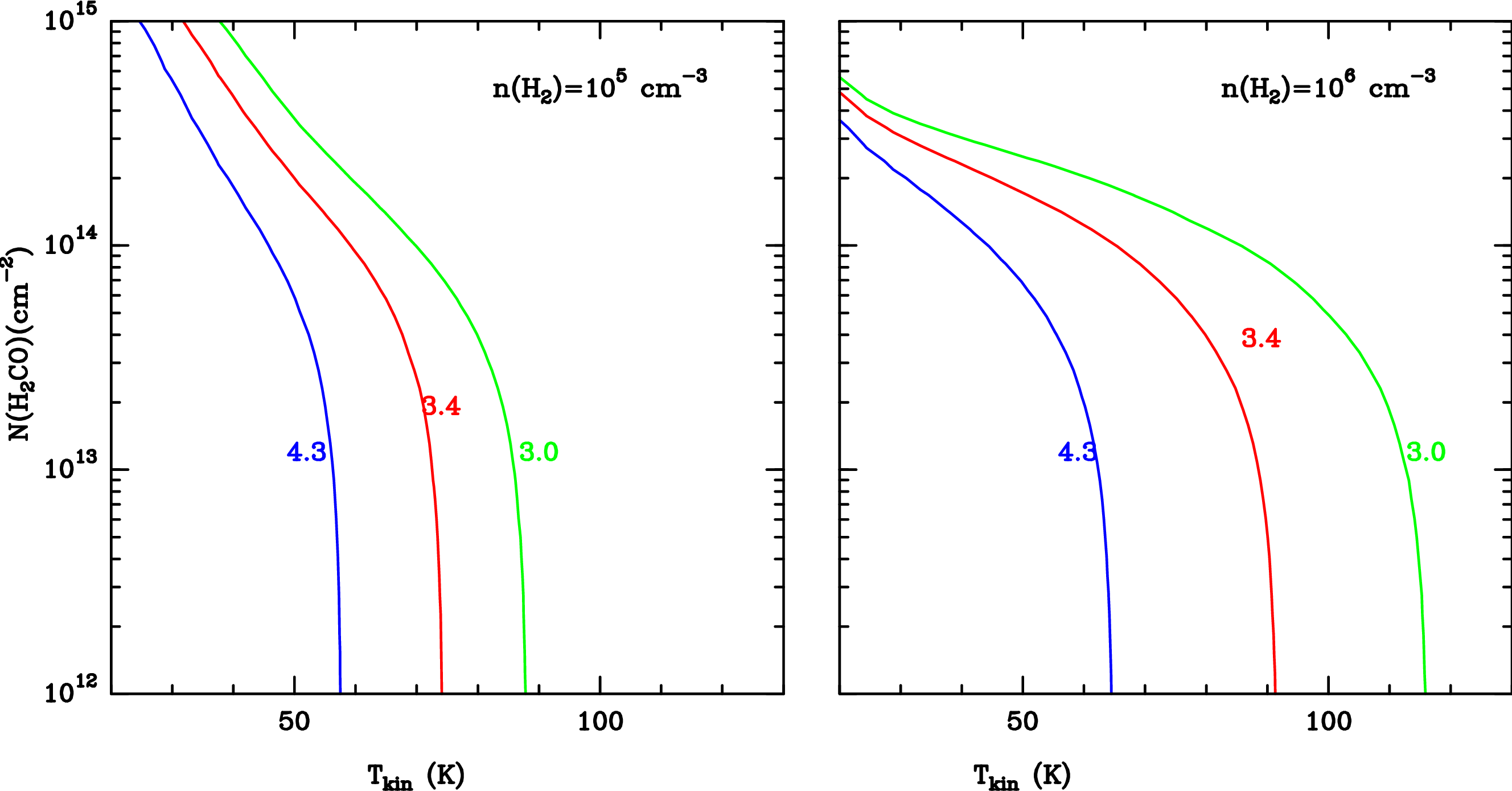}
\caption{RADEX non-LTE modeling of the
3$_{03}$--2$_{02}$/3$_{21}$--2$_{20}$ ratio of \formaldehyde\ as a
function of the gas kinetic temperature (\tkin) and column  density of
\formaldehyde\ for the volume densities $n$(H$_2$) of 10$^5$ ({\em
Left}) and 10$^{6}$\,\cmcub ({\em Right}). Contours with labels mark the 
observed line intensity ratios for \formaldehyde.
\label{fig_h2coradex}}
\end{center}
\end{figure}

The intensity ratios of the 3$_{03}$--2$_{02}$/3$_{22}$--2$_{21}$ and
the 3$_{03}$--2$_{02}$/3$_{21}$--2$_{20}$ transitions of
para-\formaldehyde\ are good thermometers for determining kinetic
temperature \citep{mangum1993}. The 3$_{21}$--2$_{20}$ has a slightly
better SNR than 3$_{22}$--2$_{21}$ in the three positions. Therefore we
use the observed 3$_{21}$--2$_{20}$/3$_{03}$--2$_{02}$ ratios to
estimate the kinetic temperature ($T_{\rm kin})$ in these positions. To
constrain the kinetic temperature, density and column density of the
gas, a grid of models based on the non-Local Thermodynamic Equilibrium
(non-LTE) radiative transfer program RADEX \citep{vdtak2007} was
generated and compared with the intensity ratios of para-\formaldehyde.
Model inputs are molecular data from the LAMDA database
\citep{schoier2005} and para-H$_2$CO collisional rate coefficients from
LAMDA \citep{wiesenfeld2013}. RADEX predicts line intensities of a given
molecule in a chosen spectral range for a given set of parameters:
kinetic temperature, column density, H$_2$ density, background
temperature and line width. A value of 2.73 K is assumed as the
background temperature for all calculations presented here.  The
synthetic line ratios are calculated for a line width of 2~\kms, close
to the line widths we deduce for the three positions. The observed line
ratios of 3$_{03}$--2$_{02}$/3$_{21}$--2$_{20}$ at the three positions
(14,26), (0,26), and (0,40) are 3.0, 3.4, and 4.4 respectively.  We
generated a grid of models with \tkin = 20--120\,K, $n$(H$_2$) =
10$^4$--10$^7$\,\cmcub\ and $N$(H$_2$CO) =
10$^{12}$--10$^{15}$\,cm$^{-2}$.  Figure\,\ref{fig_h2coradex} shows a plot
of the predicted 3$_{03}$--2$_{02}$/3$_{21}$--2$_{20}$ ratios as a
function of \tkin\ and $N$(H$_2$CO) for $n$(H$_2$) = 10$^5$ and
10$^6$\,\cmcub. The observed values of the H$_2$CO ratios are shown as
contours, which constrain the temperature very well up to $N$(H$_2$CO)
of 2$\times 10^{13}$\,\cmsq. Based on observations of high mass star forming
clumps, the value of $N$(H$_2$CO)/$N$(H$_2$) ranges between 10$^{-10}$
to 10$^{-9}$, which correspond to $N$(H$_2$CO) of
10$^{12}$--10$^{13}$\,\cmsq\ for the derived $N$(H$_2$) of $\sim
10^{22}$\,\cmsq\ for the globule.

The critical density of the \formaldehyde\ transitions that we
observe is 1--6\,10$^5$ for 50\,K gas \citep{shirley2015} and the
critical densities for low-$J$ CO and NH$_3$ are much lower ($\sim
10^3$\,\cmcub\ for NH$_3$). Thus, while  a density of 10$^4$\,\cmcub\ is
too low, $n$(H$_2$) of 10$^6$\,\cmcub\ appears to be too high. We  thus
assume a  gas density,  $n$(H$_2$) = 10$^5$\,\cmcub, which we consider
to be a reasonable compromise. The outcome of the
analysis stays about the same even for slightly higher densities. With the assumption of a gas density,
$n$(H$_2$) = 10$^5$\,\cmcub, we obtain kinetic temperatures of 55, 75,
and 90\,K respectively. However, we note that since the \formaldehyde\
lines are weak, the uncertainties in \tkin\ determined from
spectra at individual positions are also high. In order to derive a more
reliable estimate of \tkin\ we averaged the \formaldehyde\ lines over an
area where all three lines are detected. The observed line ratios then
lie in the range of 3.2--3.4 for 3$_{22}$--2$_{21}$  and
3$_{21}$--2$_{20}$, respectively. The average kinetic temperatures  is
therefore in the range 60--70 K.  \citet{breen2018} used NH$_3$ (1,1)
and (2,2), another good molecular gas thermometer \citep{wamsley1983}, to
derive a  rotational temperature, T$_{rot}$, of 23 K averaged over the
area where both (1,1) and (2,2) was detected. This correspond to a
kinetic temperature of 39 K, which is similar to what we get from
CO(2--1).  The kinetic temperatures estimated from far infrared dust
emission \citep{roccatagliata2013}, are somewhat lower.
\citet{roccatagliata2013} find $\sim$ 32 K for the head region and
$\sim$ 25 K for the column density peak. Thus, the \tkin\ derived from
\formaldehyde\ is higher than other estimates, although the reason for
this discrepancy is not clear.


\subsection{\cplus\ column density}

While \CII\ is mostly known to be optically thin, recent
velocity-resolved observations of Galactic massive star forming regions
have shown that this is not always the case, because the detection of
strong hyperfine transitions of  $^{13}$\cplus, indicate that  \CII\
can be significantly optically thick \citep[][among
others]{mookerjea2018,Ossenkopf2013}.  If \thCII\ is not detected, the
determination of \cplus\ column density, $N$(\cplus) requires
assumptions about the excitation temperature or using the same
excitation temperature as molecular emission. Simultaneous detection of
the \CII\ and \thCII\ lines is independent of calibration errors and
allow us to determine the excitation temperature and optical depth, if
\CII\ is optically thick. The F = 2--1 hyperfine line of \thCII\ was
detected in 50 positions. As described in Section~\ref{sec-Results} we
derived the integrated line intensities of the \thCII\ emission at all
these positions and below we estimate the  $T_{\rm ex}^{\rm CII}$ and
$N$(\cplus) following the formulation by \citet{Ossenkopf2013}.

The optical depth $\tau_{12}$ of the \CII\ line is determined
using 
\[
\frac{\rm I_{12}}{\rm I_{13}} = \frac{\rm ^{12}C}{\rm
^{13}C}\left[\frac{1-\exp(-\tau_{12})}{\tau_{12}}\right],
\]

where  $I_{12}$ and $I_{13}$ represent $\int T^{CII}_{\rm mb}
d\upsilon$ and  $\int T^{^{13}CII}_{\rm mb} d\upsilon$ respectively, and
 $\frac{\rm ^{12}C}{\rm ^{13}C}$ = 65 \citep{rathborne2004}.

The optical depth derived for \CII\ range between 0.9 to 5.

The optical depth-corrected integrated intensity of \CII\ is derived
from

\[
{\rm I_{12}^{\rm corr}} = \frac{\rm I_{12}\tau_{12}}{1-\exp(-\tau_{12})}
\]

Subsequently, the excitation temperature of \CII\ ($T_{\rm ex}^{\rm
CII}$) is derived by dividing Eq (1) by Eq (3) from \citet{Ossenkopf2013}

\[
\frac{\rm I_{12}^{\rm corr}}{\int{\tau_{12}{\rm d}\upsilon}} =
\frac{92\exp(-91/{\rm T_{ex}^{CII}})}{1-\exp(-91/{\rm T_{ex}^{CII}})}
\]

The values of $T_{\rm ex}^{\rm CII}$ vary between 80 to
255\,K. Most positions show values between 110--130\,K, which is
consistent with the \tkin\ derived (Sec.\,\ref{sec_tkinmol}) from H$_2$CO
and CO(11--10).

Because we know the optical depth and the excitation temperature, we
can now estimate the column density of \cplus\ by inverting the
following Equation (3) of \citet{Ossenkopf2013}:

\[
\int{\tau_{12} {\rm d}\upsilon} = 7.15\times 10^{-18} {\rm N_{C^+}} \frac{\rm
32.9~K}{\rm T_{ex}}
\]

The column density of \cplus\ ranges between 3--7$\times10^{18}$\,\cmsq.

\subsection{Mass of the cometary globule}

Since we have mapped most of the globule in CO(2--1) and isotopologues
we can use our maps to estimate the total mass of the globule. For this
we use $^{13}$CO(2--1) and compute a column density for each pixel in
the map, where $^{13}$CO(2--1) is detected. For the column density
estimates we use the LTE approximation and estimate $T_{\rm ex}$ from
the peak of  $^{12}$CO at the same position and then derive the opacity
$\tau$ for $^{13}$CO, i.e. following the method outlined in
\citet{white1995}. With the assumptions that $\frac{\rm ^{12}C}{\rm
^{13}C}$ = 65 and $\frac{\rm CO}{\rm H_2}$ = 10$^{-4}$ we get a total
mass of $\sim$ 600~\Msun.

We derive similar mass estimates from the observed \CII\ intensities by
assuming the line to be optically thin, a gas density of $n=n_{\rm
cr}$=3000\,\cmcub\ and an excitation temperature, $T_{\rm ex}$ = 100 K.
If we assume that half of the available carbon is in C$^+$ (0.65
$\times$ 10$^{-4}$), then the total mass of the globule seen in \CII\
is  $\sim$ 440 \Msun, which is comparable to what we derived from
$^{13}$CO(2--1).  This is an underestimate, since we know that  there
are positions where \CII\ is optically thick.  The fraction of pixels
where \CII\ is optically thick, however, is only $\sim$ 8\% of all
pixels in which \CII\ was detected. Therefore we at most underestimate
the mass by 15\%. The uncertainty in our mass estimate is completely
dominated by how much of the carbon is in \CII.

The masses we derive from CO and C$^+$ are also in reasonably good
agreement with the value of 1200 \Msun\ estimated from the 1.2\,mm
dust continuum emission by  \citet{rathborne2004}.  The mass estimates
from different tracers are therefore comparable and suggest a total
mass of the globule of $\sim$ 1,000 \Msun.

\subsection{Position-Velocity diagrams along the cuts}

\begin{figure}[h]
\begin{center}
\includegraphics[width=0.5\textwidth]{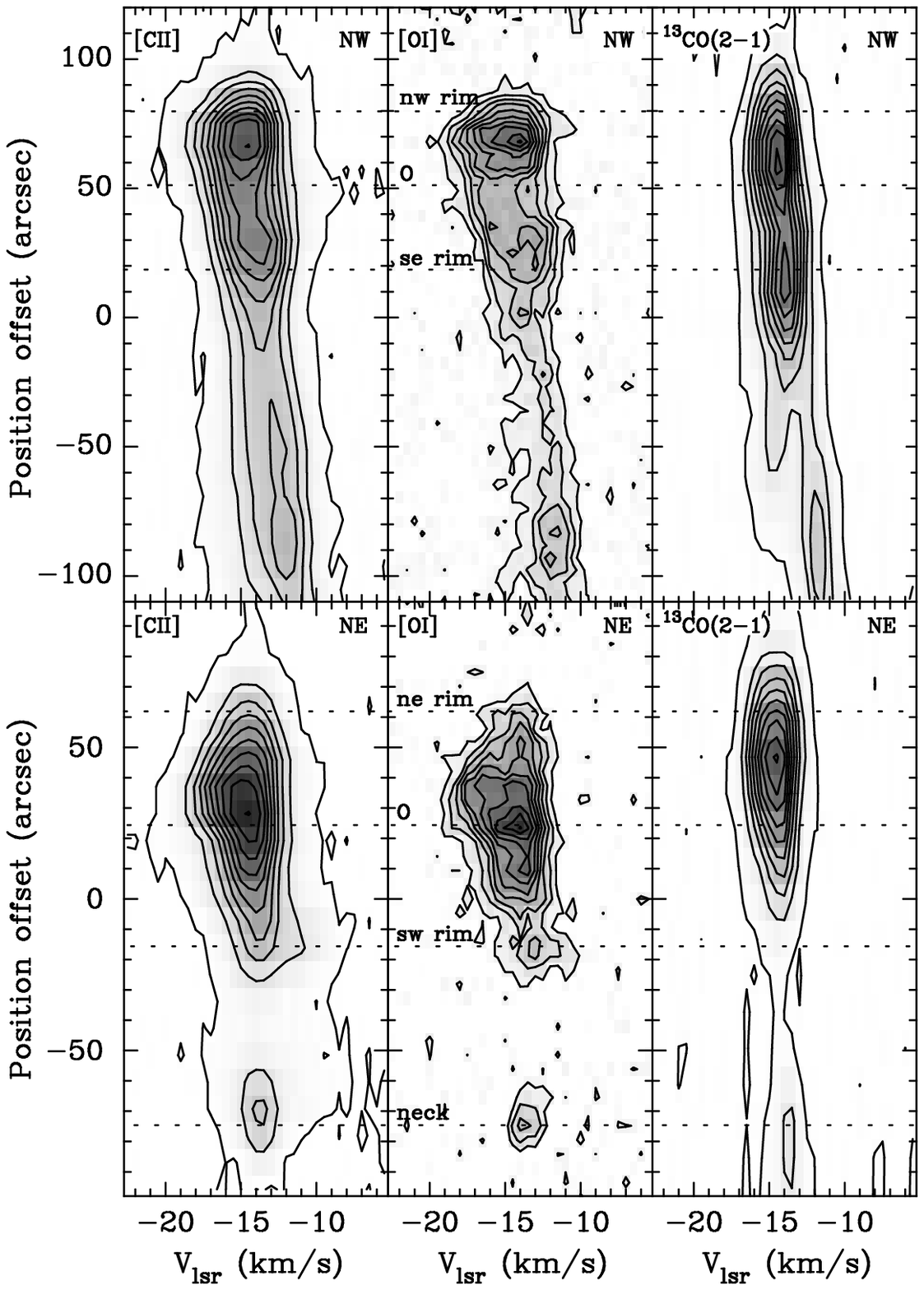}
\caption{Position-velocity plots of \CII, \OI\  and $^{13}$CO(2--1) in
contours enhanced with grayscale along the cuts shown in
Fig.~\ref{fig_g287_8um}. The zero position for the NW-SE cut (top panels) is at
(+27\arcsec,-23\ptsec5) and for the NE-SW cut (bottom panels) at
(-27\arcsec,+10\arcsec)
relative to the map center. For each cut we plot 10 contours, which for
\CII\ go from 1.4\,K to peak temperature, for \OI\ and  $^{13}$CO(2--1) lowest
contour is a 3\,K and 0.4\,K, respectively. The grayscale goes from zero
to 1.5 times the peak temperature. For all cuts we have drawn the
position of the PDR rims (as seen in 8 $\mu$m IRAC image) with a dotted
line as well as the position of the ionizing star, CPD -59\arcdeg 2661.
The rims and the O star (O) are labelled in the middle panel.  In the
NE-SW cut the dotted line labelled ``neck'' marks the middle of the
``duck'' neck.
\label{fig_pvcuts}}
\end{center}
\end{figure}

In order to understand how the PDR emission relates to the molecular gas
in the globule, we have made position-velocity (PV) plots in  \CII, \OI\ and
$^{13}$CO(2--1) along the two cuts shown in Fig.~\ref{fig_g287_8um}. The
PV plots were made using the task velplot in Miriad
\citep{sault1995}. Velplot uses a 3 $\times$ 3 pixel interpolation
kernel resulting in beam sizes of 21\ptsec2, 11\ptsec1, and  28\ptsec5
for \CII, \OI, and $^{13}$CO(2--1)\footnote{For $^{13}$CO(2--1) we
created a map with a pixel size of 9\ptsec5 $\times$ 9\ptsec5 , i.e.,
one third of a HPBW.}, respectively.  Fig.~\ref{fig_pvcuts} shows the
PV  plots along the cut going from:  (top) north-west
to south-east (NW-SE) and (bottom) north-east to south-west (NE-SW). 

Along the NW-SE cut, the PDR tracers rise sharply in the northwest and
then fall off and become narrower  after passing through the \HII\
region powered by CPD -59\arcdeg 2661.  The tail region is distinctly
red-shifted in the eastern tail of the globule, although both \CII\ and
$^{13}$CO(2--1) show a velocity component at -14.5 \kms, probably
arising from unperturbed gas from the head of the globule. As we already
saw from the channel maps, this velocity component is not seen in
C$^{18}$O or \OI, suggesting that it is more diffuse, low density gas.
The \CII\ and \OI\ PV plots show clear peaks where the cut intersects
the PDR shell illuminated by CPD -59\arcdeg 2661, while $^{13}$CO(2--1)
peaks a bit further south.  The $^{13}$CO(2--1) line width remains
unchanged across the PDR shell, which indicates that the molecular gas
in the globule is not yet affected by the expanding \HII\ region. 

In the NE-SW position-velocity plot the PDR shell stands out clearly in
\CII\ and \OI\ being more blue-shifted close to the exciting star
(marked by O in Fig.\,\ref{fig_pvcuts}) while $^{13}$CO(2--1) peaks at
the molecular column density peak.  \OI\  and \CII\ also have peaks at
the south western PDR rim of the main globule. \OI\ is not present in
the low density region between the main globule and the Duck neck, and
$^{13}$CO(2--1) is very faint or absent as well.  \CII, however, is
relatively strong and distinctly redshifted, presumably low density gas
expanding from the PDR region illuminated by CPD -59\arcdeg 2661. Both
\CII\ and \OI\ show extended emission on the side of the neck facing
towards the Treasure Chest, indicating that the neck is associated with
the PDR region illuminated by CPD -59\arcdeg 2661.  At the angular
resolution of our observations the northern PDR illuminated by
$\eta$\,Car and  the PDR shell illuminated by CPD -59\arcdeg 2661 blend
together, whereas they are seen as two separate PDRs in the 8 $\mu$m
image, see Fig.~\ref{fig_g287_8um}. The northern PDR, however, is close
to the systemic velocity of the globule head, while the internal PDR
shell is more blue-shifted north of the star. In the NE-SW cut we see a
similar jump in velocity between the external PDR rim and the PDR shell.  

Both position-velocity plots show that there is \CII\ emission
outside the thin PDR interface of the globule. This is the same photo
evaporated gas seen for example in ionized Paschen $\beta$
\citep{hartigan2015}. At the sharp northern PDR surface, one can see
faint \CII\ emission tracing photo evaporated gas up to $\sim$ 20\arcsec\
north of the head of the globule.

\section{Discussion}

\begin{figure}[h]
\begin{center}
\includegraphics[width=0.25\textwidth]{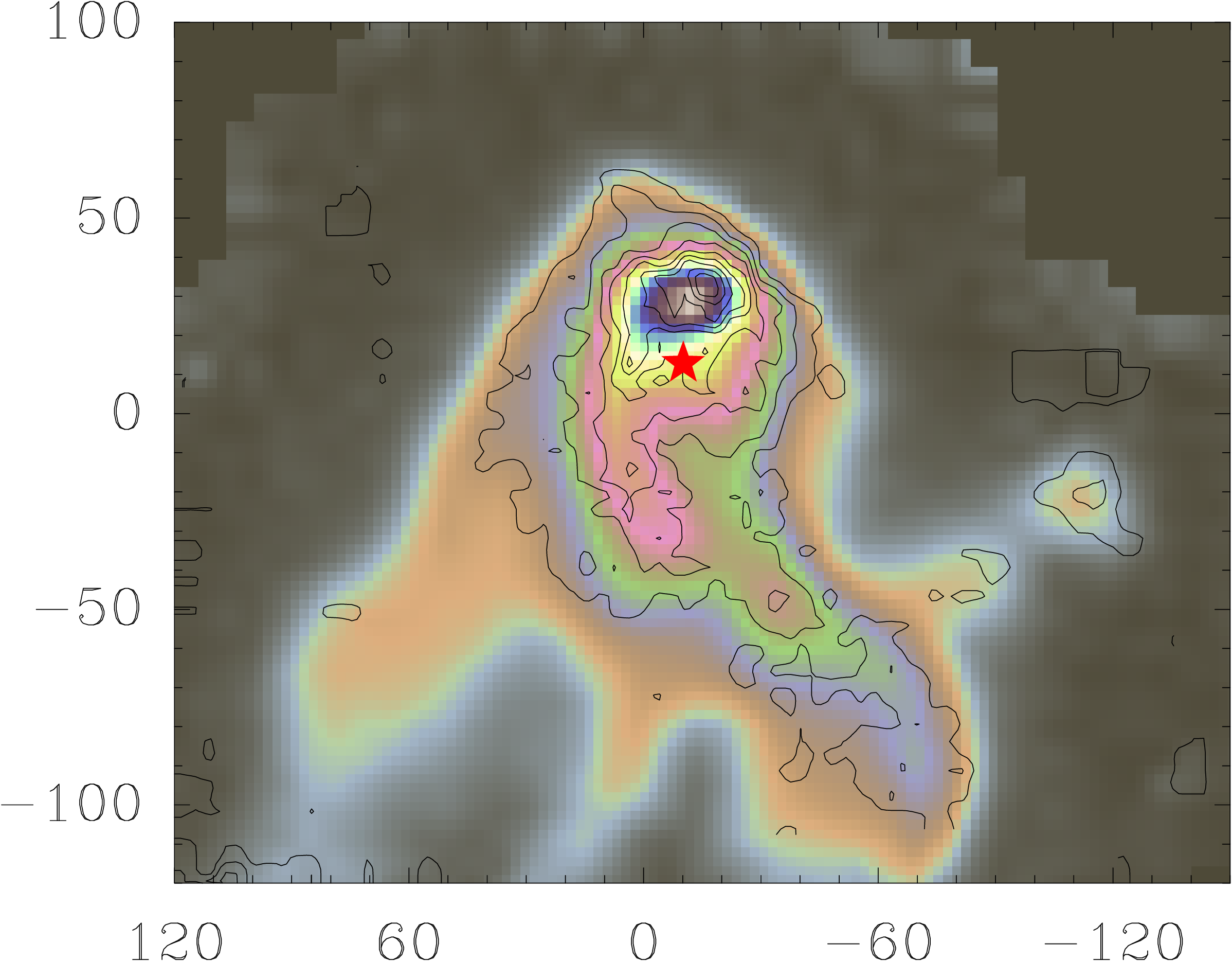}
\includegraphics[width=0.225\textwidth]{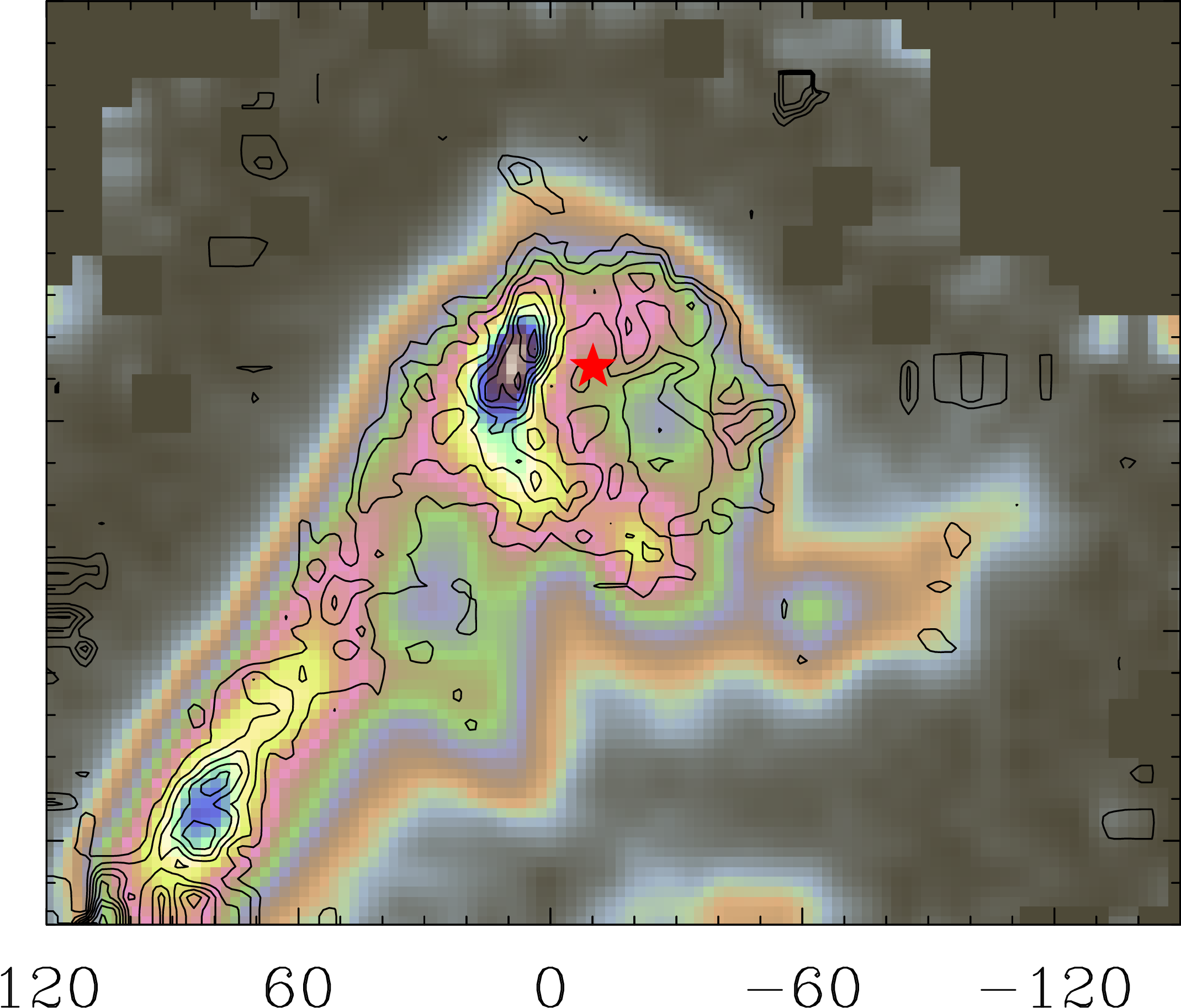}
\caption{Comparison of integrated line intensities of \CII\ (color) and
\OI\ at 63\,\micron\ (contour) over different velocity ranges. {\em
Left:}
Blue velocity component between -19 to -14\,\kms\ and {\em Right:} Red
velocity component between -13 to -10\,\kms.  The filled red asterisk shows
the position of CPD-59\arcdeg 2661.  The positional offsets are relative 
to the center $\alpha$=10$^h$45$^m$55\fs05,$\delta$=-59$^d$57$^m$16\farcs7 
(J2000).
\label{fig_OIcp}}
\end{center}
\end{figure}

We compare the high spatial resolution (7\arcsec) velocity-resolved
observations of the \OI\ at 63\,\micron\ lines in G287.84-0.82 with the
\CII, CO (and its isotopologues) and \formaldehyde\ to understand the
small-scale structure of the PDR created by the embedded Treasure Chest
cluster. As shown in Fig.\,\ref{fig_chanmaps} the emission from the
region has clearly blue- and red-shifted components.
Figure\,\ref{fig_OIcp} compares the emission in the blue ($\upsilon_{\rm
LSR}$ = -19 to -14\,\kms) and red ($\upsilon_{\rm LSR}$ = -13 to
-10\,\kms) parts of the spectrum. In the blue part we find \OI\ tracing
out mostly the higher density regions including the PDR surface to the
southwest of the main pillar. The same is seen in \CII, which
additionally is seen in the tail regions of the globule.  The overlay of
red-shifted \CII\ and \OI\ emission also match very well, suggesting the
PDR shell surrounding  CPD -59\arcdeg 2661 is dominantly at this
velocity. The eastern, red-shifted tail of the globule is strong in both
\CII\ and \OI, while the middle tail is only seen in \CII, suggesting
that it has lower densities.  The western pillar (the Duck Head) clearly
stands out in the high spatial resolution \OI\ data and is seen in \CII\
as well. It appears to have about the same velocity as the main pillar.
The head is seen in CO(11--10) as well.

The higher spatial resolution \OI\ data resolve the PDR shell around
the Treasure Chest cluster, with the semi shell being easily seen in
the red-shifted \OI\ image. This is also consistent with our \CII\
data. There is little overlap between the peaks in the blue- and
red-shifted emission, which is consistent with an expanding shell, in
which the north-eastern part is somewhat blue-shifted, while the
emission is more red-shifted to the west. The red- and the blue-shifted
components of \CII\ and \OI\ emission differ in velocity by $\sim$
4~\kms. This is  much smaller than the velocity split seen in
H$\alpha$, 25 \kms, toward CPD -59\arcdeg 2661 by \citet{walsh1984}.
Although \citet{walsh1984} interpreted it as a measure of the expansion
velocity (12 \kms{}) of the nebula, he noted that "there is no evidence
for the split components gradually joining up to the velocity of the
adjacent material, as might be expected for a shell." Our observations
show that the expansion velocity is only $\sim$ 2 \kms, hence the
dynamical age of the \HII\ region, a few times 10$^4$ yr  estimated by
\citet{smith2005}, is severely underestimated.  In order for the \HII\
region to expand into the surrounding dense molecular cloud, a more
realistic estimate of the age is $\sim$ 1 Myr and this is in good
agreement with the age of the cluster, 1.3 Myr, as estimated by
\citet{oliveira2018}.

\subsection{The PDR gas}

We can to some degree distinguish the emission from the diffuse and
dense components of the gas by looking at the emission in different
velocity intervals. Thus, if we compare the intensities due to the
blue-shifted \CII\ and CO(2--1) emission from the eastern tail with PDR
models, it will be possible to derive an estimate of the density of the
diffuse gas.  Additionally, we use the line intensities of \OI\ and
CO(11--10), both high density PDR tracers, at the positions of the
intensity peaks of \OI\ seen clearly in the shell around the embedded
cluster (Fig.\,\ref{fig_OIcp}). 

In order to estimate the physical conditions in the  dense and diffuse
PDR in the cloud surrounding the Treasure Chest nebula  we compare the
observed line intensity ratios with the results of the model for PDRs
by \citet{Kaufman06}.  These models consider a semi-infinite slab of
constant density, which is illuminated by far-ultraviolet (FUV) photons
from one side.  The model includes the major heating and cooling
processes and incorporates a detailed chemical network. Comparison of
the observed intensities with the steady-state solutions of the model,
helps to determine the gas density of H nuclei, $n_{\rm H}$, and  the FUV
flux (6~eV $\leq$ h$\nu$ $<$ 13.6~eV), $G_0$, measured in units of the
\citet{Habing1968} value for the average solar neighborhood FUV flux,
1.6$\times 10^{-3}$ ergs\,cm$^{-2}$\,s$^{-1}$. 

Since the PDR inside the Treasure Chest cloud is  created by the
embedded cluster, the strength of the FUV radiation in the region can
be estimated from the observed total far-infrared (FIR) intensity, by
assuming that FUV energy absorbed by the grains is re-radiated in the
FIR. Based on this method \citet{roccatagliata2013} have derived the
FUV map for the region using PACS and SPIRE dust continuum
observations, with the values of FUV radiation field lying between 900
to 5000 times G$_0$, where G$_0$ is in units of the Habing field.  From
their FUV map (Fig.\,7 in \citet{roccatagliata2013}) we estimate the
FUV intensity to be around 1000\,G$_0$ in the eastern tail and around
5000\,G$_0$ in the region where the PDR shell is detected.

\begin{figure}[h]
\begin{center}
\includegraphics[width=0.25\textwidth]{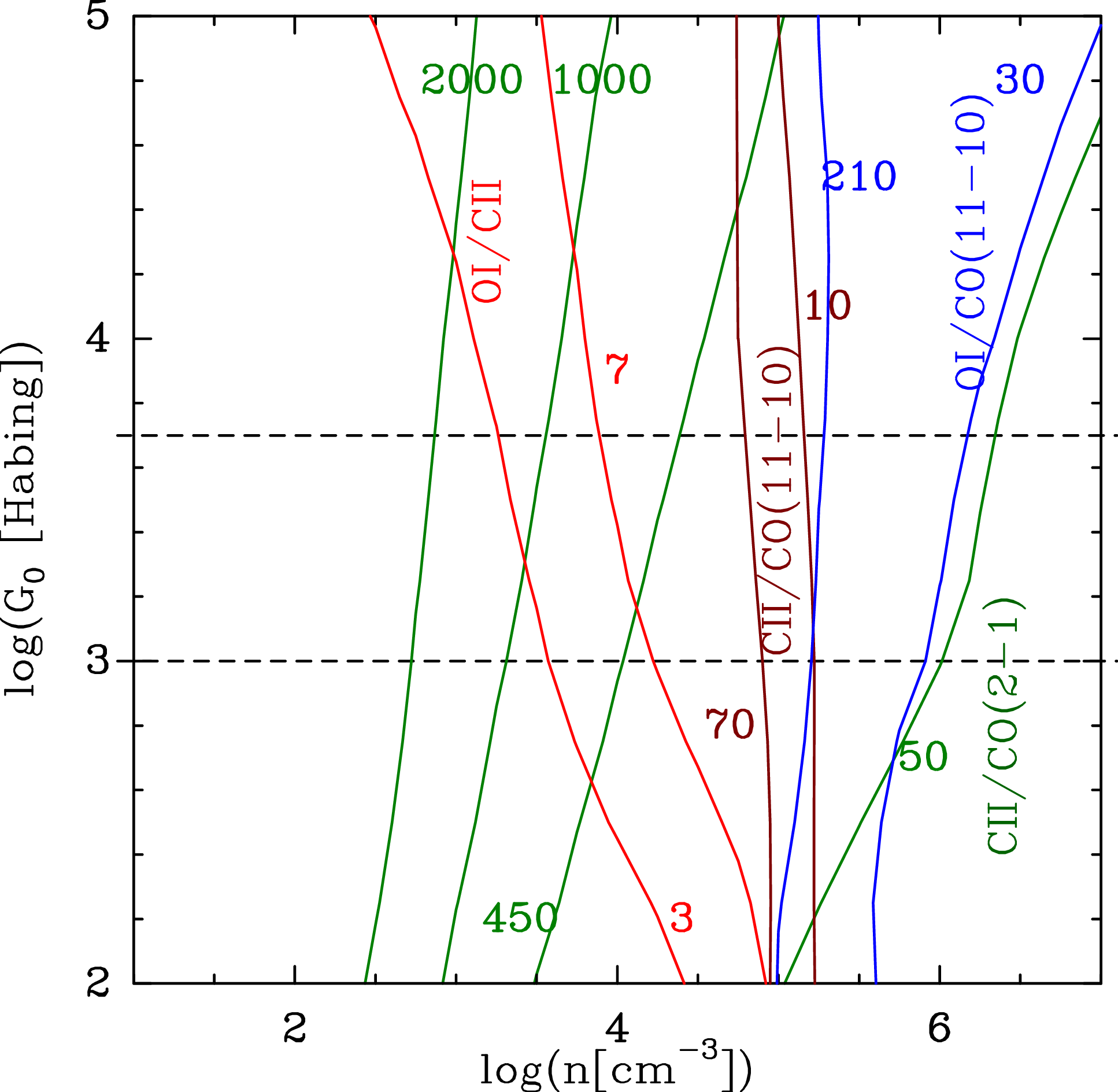}
\caption{Contours of observed intensity ratios plotted on the
intensity predictions as a function of hydrogen density $n_{\rm H}$ and
FUV radiation field ($G_0$) from PDR models, for the selected regions in
the cloud around Treasure Chest. The horizontal dashed lines correspond to 
FUV in the eastern tail (1000\,G$_0$) and close to the embedded cluster
(5000\,G$_0$) as estimated by \citet{roccatagliata2013}. The contours
of \CII/CO(2--1) (green), \CII/CO(11--10) (brown), \OI/CO(11--10)
(blue) and \OI/\CII\ (red)
correspond to the extreme values that these ratios (see text) take up.
The ratios are in energy units.
\label{fig_pdrmodel}}
\end{center}
\end{figure}

In order compare the observed intensity ratios of \CII/CO(11--10),
\OI/CO(11--10), \OI/\CII\ and \CII/CO(2--1) for the dense gas with PDR models,
we compare emission arising in the same velocity range for the tracers involved
in a particular ratio.  Since among all the tracers \CII\ is the broadest at
most positions, we first fit a single Gaussian to the spectrum of the other
tracer in the ratio.  Subsequently we fit a Gaussian with identical central
velocity and FWHM to obtain the velocity integrated line intensity for \CII. In
order to do this position-by-position analysis we spatially regridded the
datacubes corresponding to a particular ratio to the grid of the cube with
lower resolution. Thus the analysis is restricted to around 300 positions where
such fits were possible.  We find that the intensity ratios (in energy units)
\CII/CO(11--10), \CII/CO(2--1) and \OI/\CII\ range between 10--70, 50--450, and
3--7 respectively for these positions. For the \OI/CO(11--10) we fit the
CO(11--10) spectra first and then use the same parameters to extract the
intensity of the \OI\ spectrum. The \OI/CO(11--10) intensity ratio thus
estimated range between 30 to 210.  For the diffuse gas traced only by \CII\
and CO(2--1), we use the ratio of intensities of  \CII\ and CO(2--1) between
-18 to -14\,\kms\ in the eastern tail region, where no \OI\ emission is seen at
these velocities.  The CO(2--1) intensities are estimated from the observed
intensities of the optically thin $^{13}$CO(2--1) spectra. The \CII/CO(2--1)
ratio for the eastern tail ranges between 1000--2000 (in energy units).


Figure\,\ref{fig_pdrmodel} shows results of comparison of the observed
intensity ratios with the predictions of the PDR models. The two
horizontal lines correspond to 1000 and 5000\,G$_0$ for the western
tail and the PDR shell around the cluster. For each intensity ratio two
contours are drawn to mark the range of observed values.  We assume
that the beam-filling factors of the different PDR tracers are
identical so that the line intensity ratios considered here are
independent of the beam-filling.  

The observed \CII/CO(2--1) ratios for the diffuse gas in the eastern
tail are consistent with $n$(H) between 600 to 2200\,\cmcub. For the
denser gas the \CII/CO(2--1) suggest a somewhat wider range of values
between 10$^4$--10$^6$\,\cmcub. The \CII/CO(11--10) primarily estimated
in the head of the globule suggest a narrower range of densities
between (0.8--1.6)$\times 10^{5}$\,\cmcub. The \OI/CO(11--10) ratio
with both high density tracers suggest densities between (2--8)$\times
10^5$\,\cmcub.  The \OI/\CII/ ratio estimated for positions spanning
both the head and tails of the globule suggest densities between
2\,10$^3$--1.8\,10$^4$\,\cmcub. The much lower densities suggested by
the \OI/\CII/ ratio indicate that the \CII\ emission is dominated by
lower density, more diffuse PDR gas.

\subsection{Thermal Pressure in the region}

Based on the analysis of the tracers of molecular and PDR gas of
different densities we identify several temperature components in the
G287.84-0.82 globule.  The first component corresponds to the
temperature of the molecular gas, which is estimated to be $\sim 40$\,K
based on the observed peak temperatures of CO(2--1).  The second
component corresponds to PDR gas at a temperature of 100\,K, as derived
from the analysis of \thCII\ intensities.  Additionally, the high
density ($\sim 10^5$\,\cmcub) molecular gas emitting in \formaldehyde\
show average kinetic temperatures of 60--70\,K within a limited region
in the head of the globule. It is likely that a fourth component with
densities of $\sim 600$--2200\,\cmcub, and not so well determined
temperature exist in the tail regions. 

From analysis of \OI\ and CO(11--10) using PDR models we determine the
densities of dense PDR gas to lie between (2--8)$\times 10^5$\,\cmcub.
For typical PDR temperatures of 100-300\,K, this corresponds to a
thermal pressure of 2\,10$^7$--2.4\,10$^8$ K\,\kms. Ratio of
intensities involving \CII\ suggest for the diffuse PDR gas, densities
around 10$^4$\,\cmcub, which for a temperature of 100\,K correspond to
a thermal pressure of 10$^6$\,\cmcub. The global pressure of the
molecular gas traced by CO(2--1) with a density of $\sim 10^4$\,\cmcub\
and \tkin = 40\,K corresponds to 4$\times 10^5$\,K\,\cmcub. If we
consider the diffuse PDR gas in the eastern tails with densities
between 600--2200\,\cmcub\ to have thermal pressure ($P_{\rm th}$) of
the same order as the cold molecular gas, then the temperature of the
gas would correspond to temperatures exceeding 200\,K.

The sharp contrast in estimated thermal pressure of the diffuse and dense phases
of PDR gas in the region, is similar to the results of \citet{stock2015}, where
two phases with pressures of 10$^5$\,K\,\cmcub\ and 10$^8$\,K\,\cmcub were
detected in multiple sources. It is also similar to the results of
\citet{wu2018}, who estimate $P_{\rm th}\sim 10^8$\,K\,\cmcub\ across
dissociation front in Car I-E region.  \citet{wu2018} has also suggested an
empirical relationship $P_{\rm th}=2.1\times10^4\,G_{\rm UV}^{0.9}$ between the
UV field and thermal pressure. For G$_{\rm UV}$ $\sim$ 5000 G$_0$ for the dense
regions, $P_{\rm th}$ will be $\sim 4\times 10^7$\,K\,\cmcub, which is
approximately consistent with the $P_{\rm th}$ we derive from the $n$(H$_2$) and
\tkin.

\subsection{Is the Treasure Chest a case of triggered star formation?}

Although it is clear that star formation can be triggered in cometary globules
\citep{lefloch1997,ikeda2008,mookerjea2009}, this is not the case for the
Treasure Chest cluster. There are 19 OB clusters in the Carina Nebula
\citep{tapia2003,smith2006,oliveira2018}.  Even though the Treasure Chest is one
of the youngest, 1.3 Myr \citep{oliveira2018}, it is still too old to have been
triggered by the formation of the cometary globule. It is far more likely that
the Treasure Chest had already formed before G287.84 became a cometary globule
and then it was a dense, massive cloud core surrounded by lower density gas.
When the Carina \HII\ region expanded, it blew away the lower density gas, first
creating a giant pillar, which later became the cometary globule that we see
today.  Although the cometary globule continues to be eroded by the stellar
winds from $\eta$\,Car and Trumpler 16, the expanding \HII\ region from the
Treasure Chest cluster appears to erode it even faster. It has already expanded
through the western side of the globule and most of the molecular cloud is
already gone on the northwestern side as well. There is, however enough dense
gas, $\sim$ 1,000 \Msun, which could collapse and form stars, although it is
likely that the time scale for star formation is longer than the time it takes
for the globule to be evaporated by the FUV radiation from the Treasure Chest
cluster and the FUV radiation from $\eta$\,Car and and the O stars in Trumpler
16.

 The heavily reddened young star, 2MASS J10453974-5957341 in the Duck head,
 could be a case of triggered star formation, although it is more likely that it
 formed about the same time the Treasure Chest cluster formed, and is therefore
 more likely a member of the Treasure Chest cluster.

\section{Summary}

We have used velocity-resolved observations of various tracers of (i)
PDR (\OI, \CII, CO(11--10)), (ii) ambient molecular material ($J$=2--1
transitions of CO, \thCO\ and \CeiO), and (iii) high density gas
(H$_2$CO and SO) to study detailed physical and velocity structure of
the G287.84-0.82 cometary globule in Carina.  The high quality of \CII\
data enabled simultaneous detection of the $F$=2--1 transition of the
rarer isotope of \cplus\ and subsequent determination of
$N$(\cplus)$\sim 0.3$--1$\times 10^{19}$\,\cmsq.  We conclude that the
overall structure of the source including its velocities is consistent
with the globule being sculpted by the radiation and winds of the nearby
OB associations in Trumpler 16 and $\eta$\,Carina. The details of the
distribution of PDR gas inside the globule as revealed with enhanced
clarity by the higher spatial resolution (7\arcsec) \OI\ data, is
consistent with illumination by the brightest member, CPD -59\arcdeg
2661 of the Treasure chest cluster.  The density and temperature of the
PDR gas, estimated to consist of a diffuse and a dense part were
determined using observed line intensity ratios in combination with
non-LTE radiative transfer models as well as plane-parallel PDR models.
We identify at least two components of PDR gas with densities of 10$^4$
and (2--8)$\times10^5$\,\cmcub.  Based on temperatures estimated, the
thermal pressure in the diffuse and dense PDR components are
10$^6$\,K\,\cmcub\ and 10$^7$--10$^8$\,K\,\cmcub\ respectively. This
contrast in thermal pressure of the diffuse and dense phases of PDR gas
is similar to the findings which exist in the literature.


\begin{acknowledgements}
 Based in part on observations made with the NASA/DLR 
Stratospheric Observatory for Infrared Astronomy (SOFIA). 
 SOFIA is jointly operated by the Universities Space Research Association, Inc. (USRA), 
under NASA contract NNA17BF53C, and the Deutsches SOFIA Institut (DSI) under 
 DLR contract 50 OK 0901 to the University of Stuttgart. 
\end{acknowledgements}

\newpage
\begin{appendix}

\section{Additional channel maps}

Here we show channel maps of CO(11--10), 2--1 transitions of CO (and
isotopologues), \formaldehyde(3$_{03}$--2$_{02}$) and
SO(6$_5\rightarrow5_4$).

\begin{figure}[h]
\begin{center}
\includegraphics[width=0.45\textwidth]{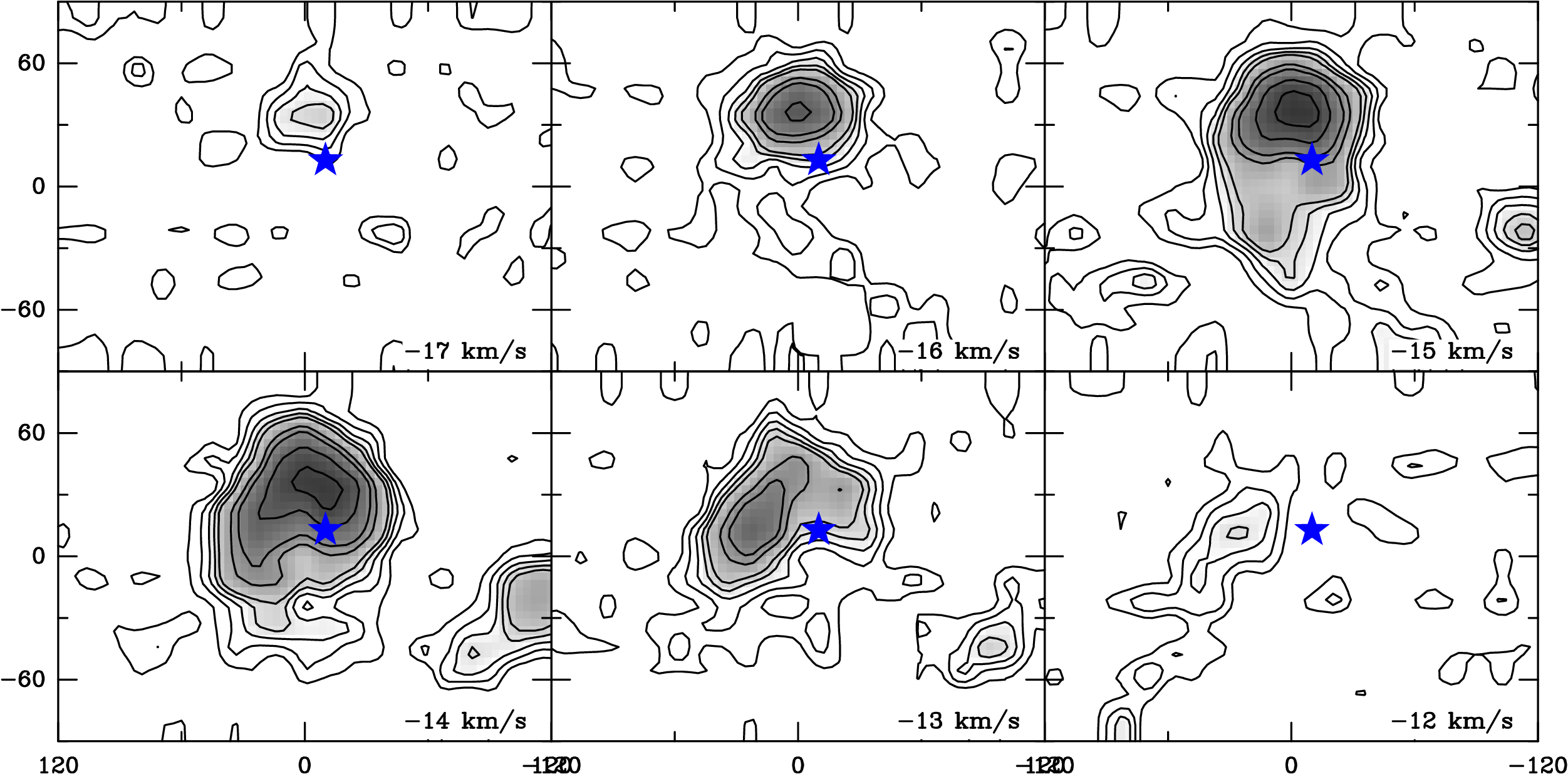}
\caption{Velocity channel maps of CO(11--10) of G287.84.
\label{fig_co11chan}}
\end{center}
\end{figure}

\begin{figure}[h]
\begin{center}
\includegraphics[width=0.4\textwidth]{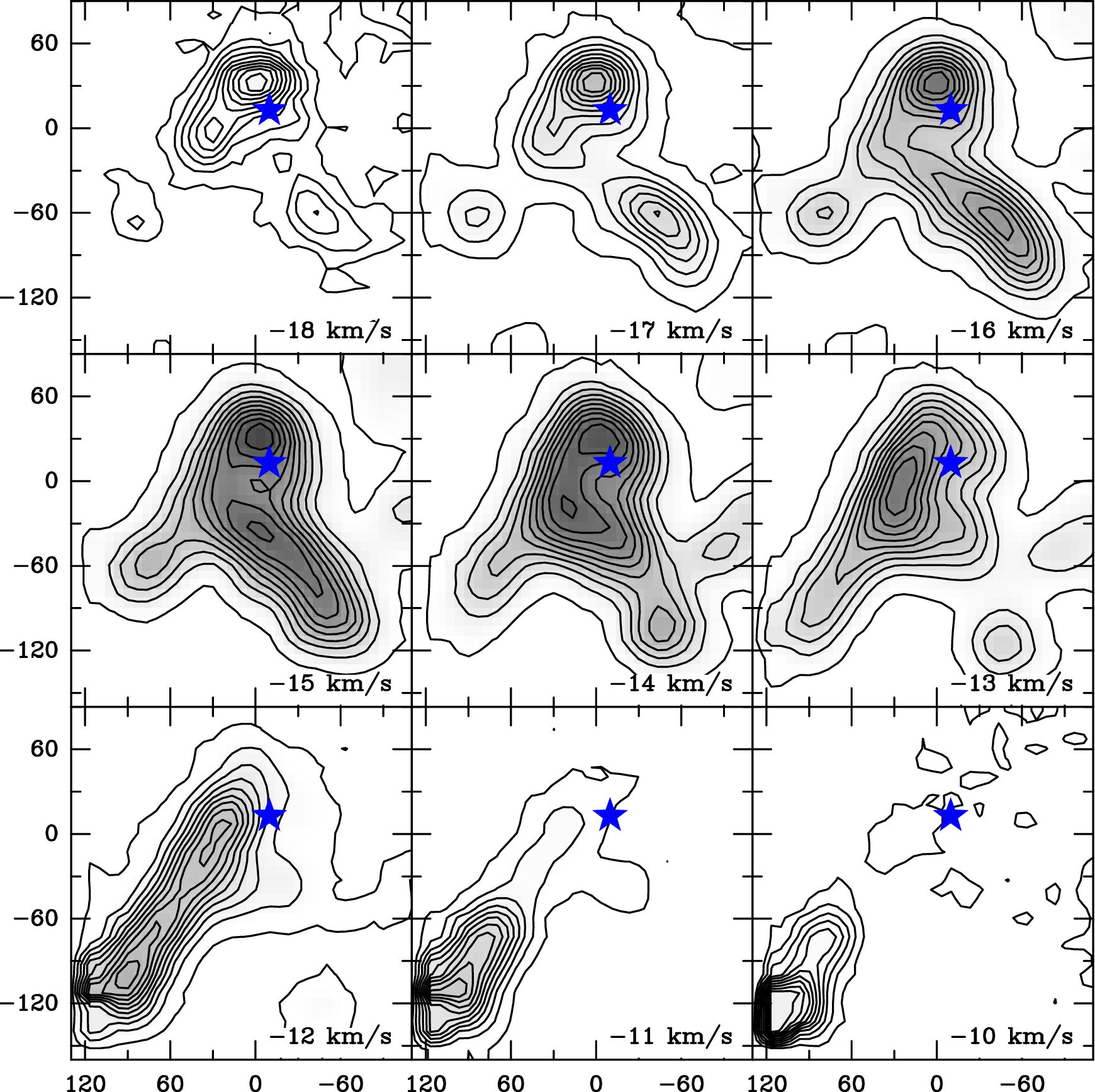}
\caption{Velocity channel maps of CO(2--1) G287.84.
\label{fig_co21chan}}
\end{center}
\end{figure}

\begin{figure}[h]
\begin{center}
\includegraphics[width=0.4\textwidth]{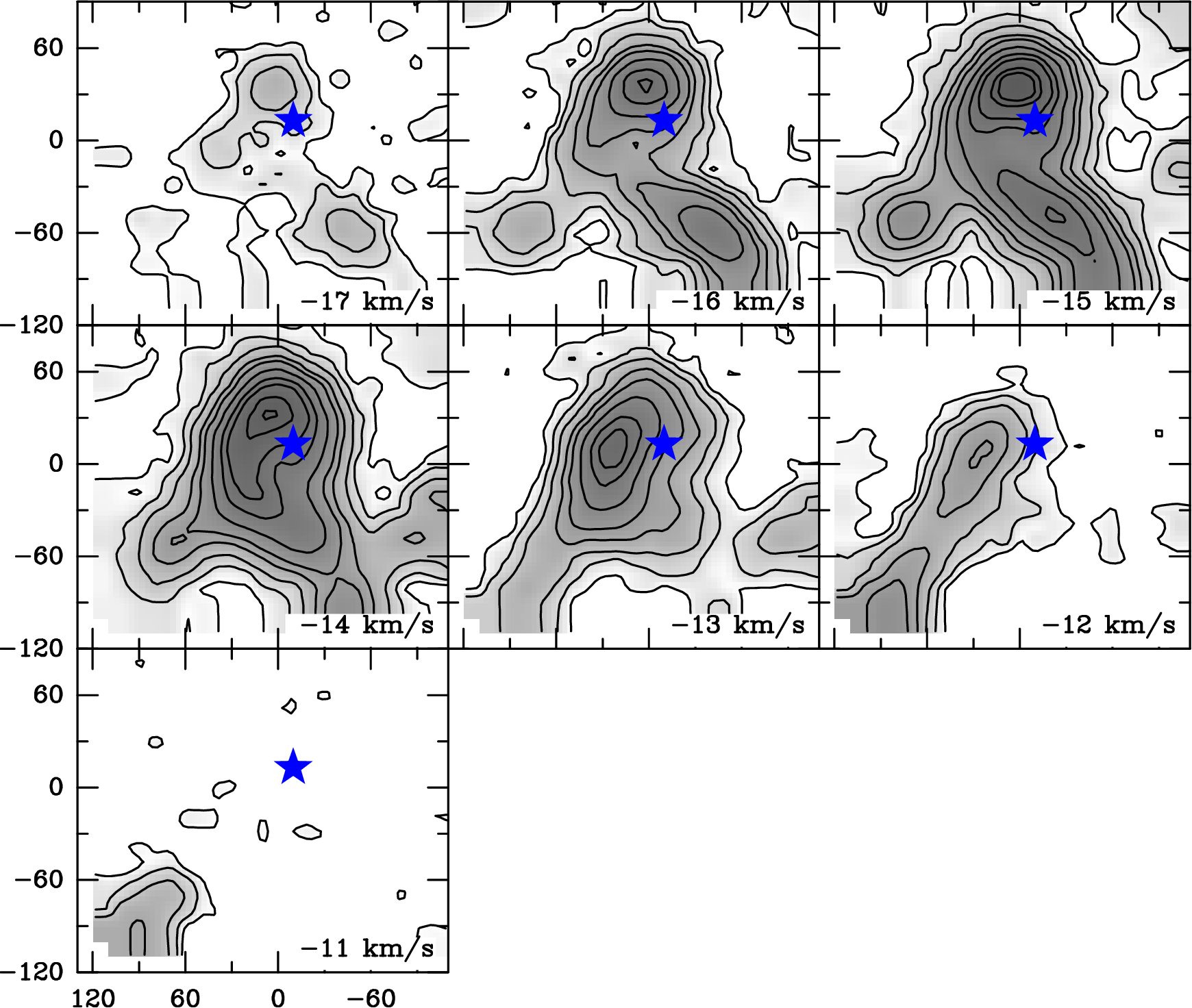}
\caption{Velocity channel maps of $^{13}$CO(2--1)  of G287.84.
\label{fig_13cochan}}
\end{center}
\end{figure}

\begin{figure}[h]
\begin{center}
\includegraphics[width=0.4\textwidth]{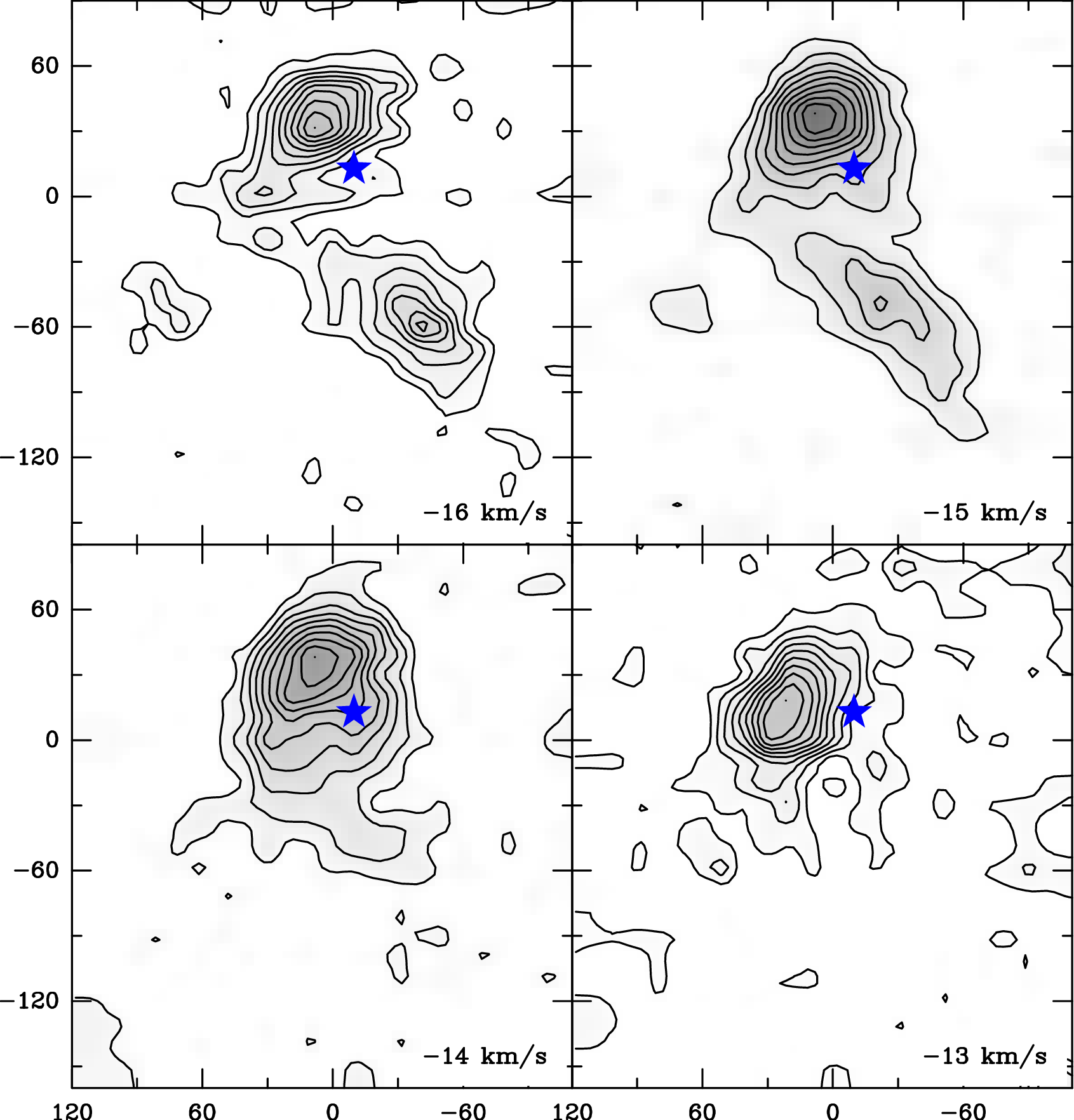}
\caption{Velocity channel maps of C$^{18}$O(2--1)  of G287.84.
\label{fig_c18ochan}}
\end{center}
\end{figure}

\begin{figure}[h]
\begin{center}
\includegraphics[width=0.4\textwidth]{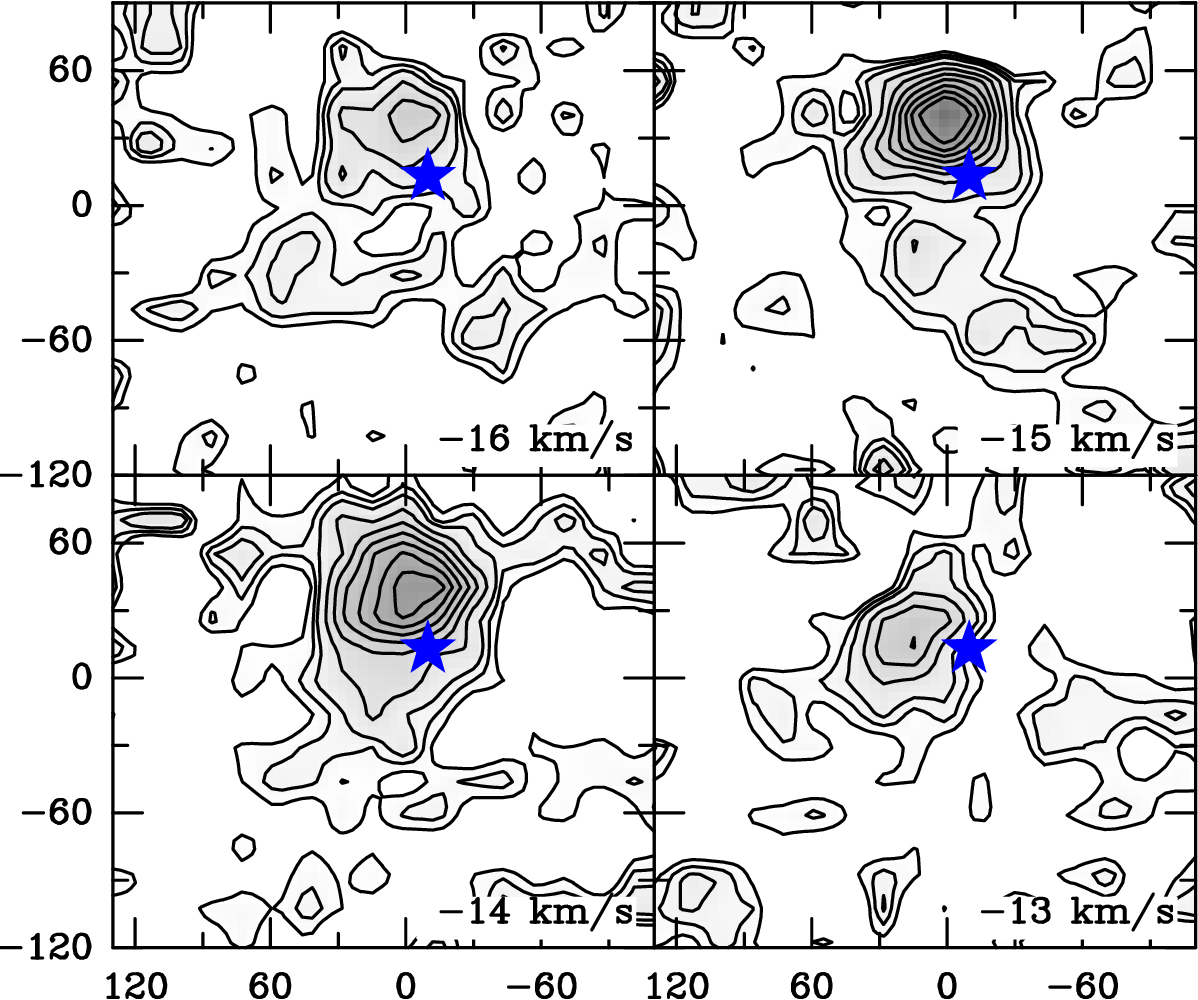}
\caption{Velocity channel maps of \formaldehyde(3$_{03}$--2$_{02}$).
\label{fig_h2cochan}}
\end{center}
\end{figure}

\begin{figure}[h]
\begin{center}
\includegraphics[width=0.5\textwidth]{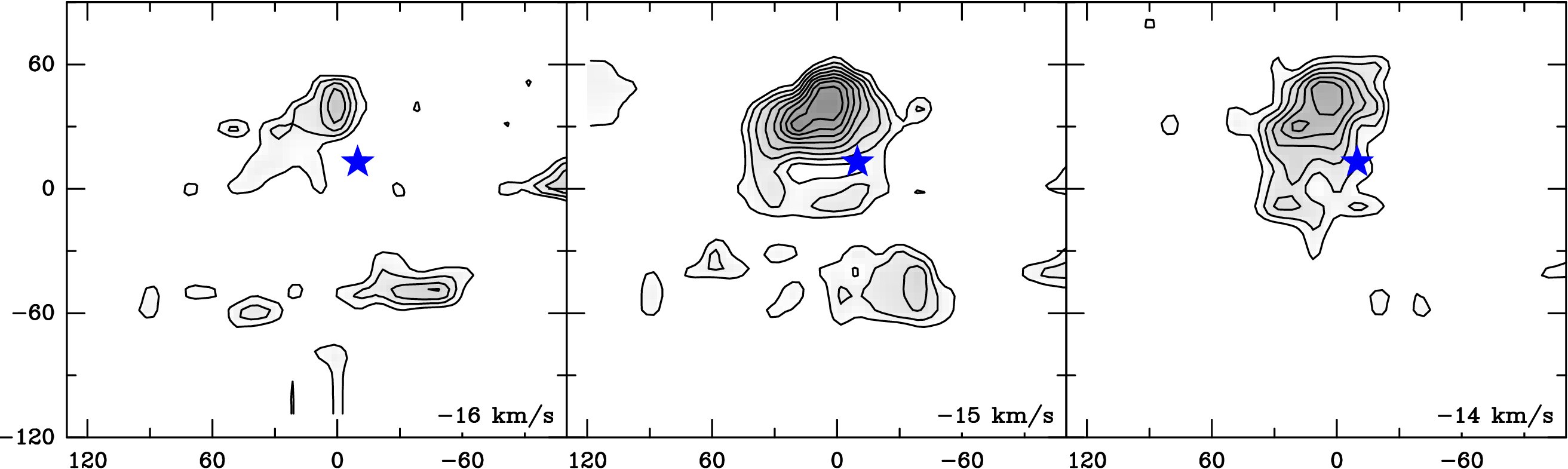}
\caption{Velocity channel maps of SO(6$_5\rightarrow5_4$).
\label{fig_sochan}}
\end{center}
\end{figure}

\end{appendix}

\end{document}